\documentclass[aps,pre,showpacs,twocolumn,floatfix,nofootinbib]{revtex4}
\usepackage{graphicx}
\usepackage{color}
\newtheorem{theorem}{Theorem}
\def\beq{\begin{eqnarray}}
\def\eeq{\end{eqnarray}}

\begin{document}

\title{Spectroscopy of drums and quantum billiards: perturbative and non-perturbative results}
\author{Paolo Amore}\email{paolo.amore@gmail.com} 
\affiliation{Facultad de Ciencias, CUICBAS, Universidad de Colima, \\
Bernal D\'{\i}az del Castillo 340, Colima, Colima, Mexico}

\begin{abstract}
We develop powerful numerical and analytical techniques for the solution of the Helmholtz equation on general domains. 
We prove two theorems: the first theorem provides an exact formula for the ground state of an arbirtrary membrane, 
while the second theorem generalizes this result to any  excited state of the membrane. 
We also develop a systematic perturbative scheme which can be used to study the small deformations of a 
membrane of circular or square shapes. We discuss several applications, obtaining numerical and analytical results. 
\end{abstract}
\maketitle

\section{Introduction}
\label{intro}

The calculation of the frequencies of a drum of arbitrary shape is a problem of formidable
difficulty. In fact, the exact solutions to the Helmholtz equation are known only in special 
cases, namely for the equilateral triangle, for rectangles and for circles and ellipses, 
where they are expressed in terms of trigonometric, Bessel and Mathieu functions.
For general shapes, including those obtained performing small perturbations of the domains just
mentioned, exact results are not available. In these cases one is forced to use approximate 
methods, which can be either analytical or numerical.

The same problem holds for a quantum billiard, which is defined as a region of space where a point
particle (for example an electron) is moving freely, not being allowed to escape. This is the quantum 
analog of a classical billiard where a particle only interacts when colliding elastically with 
the walls of the billiard, changing its trajectory. The Schr\"odinger equation in this case reduces 
to the Helmholtz equation used to describe the drum. 

The study of quantum billiards requires the calculation of a large number of energy levels, which 
are needed to extract the statistical distribution of the the level spacing $s_n = E_{n+1}-E_{n}$.
On theoretical grounds it is expected  to obtain different statistical distributions if 
the corresponding classical billiard is either integrable or chaotic.

Given the interest in calculating both the low and high parts of the spectrum, different approaches 
have been developed over the years to obtain approximate solutions to the general problem. 
We may divide them into two categories: numerical and analytical
methods. It is outside of the scope of this paper to give a full account of all different approaches 
which have been used before for this problem; we rather prefer to mention a selected 
number of works which we find representative and which can the source for additional reading.
The reader interested in finding more details and references may find useful to check the bibliography
of the papers that we mention; in particular, a very useful source of information on this problem is the
beautiful paper by Kuttler and Sigillito ~\cite{Kuttler84}, which contains $142$ references 
and which reports both numerical and analytical results.

Among the numerical methods used to solve the Helmholtz equation we mention:
\begin{itemize}
\item the method of particular solutions (also known as Point Matching Method) of Fox, Henrici and Moler ~\cite{Fox67} (a recent
      improvement of this method is discussed in ref. \cite{Betcke05});
\item the method of fundamental solutions of Bogomolny ~\cite{Bogomolny85};
\item domain decomposition method of ref.~\cite{Descloux83} (a recent application to isospectral drum is contained in \cite{Driscoll97});
\item the conformal mapping method (CMM) of Robnik~\cite{Robnik84};
\item the plane wave decomposition method (PWDM) of Heller ~\cite{Heller91};
\item boundary integral methods (BIM) (see for example ~\cite{Kostin97,Heller04});
\item the method of Vergini and Saraceno ~\cite{Vergini95};
\item the expansion method of Kaufman, Kostin and Schulten \cite{Kaufman99};
\item finite element methods (FEM), see for example \cite{DeMenezes07, Son09} or the multigrid method of 
      Heuveline \cite{Heuveline03};
\item collocation methods (in particular ref.~\cite{Boyd00} contains a useful introduction to 
      pseudospectral methods): as a special case in this category we mention the conformal 
      collocation method \cite{Amore08, Amore09}, which will be used in this paper;
\item the radial rescaling method of  \cite{Lijnen08}
\end{itemize}

We now come to the analytical approaches to the solution of the Helmholtz equation, which are essentially based 
on perturbation theory. It appears in this case that the attempts to perform systematic and precise calculations
for the spectrum of arbitrary membranes are fewer and not as successfull as for the numerical methods.
In particular Ref.~\cite{Feshbach53} describes a boundary shape perturbation method, which is 
expressed in terms of the Green's function. Ref.~\cite{Walecka80} describes the arbitrary perturbation
of a circular domain which preserves the internal area of a membrane. This first order calculation is 
originally due to Rayleigh~\cite{Rayleigh45}. Other perturbative approaches have been
introduced in \cite{Parker98, Read96, Bera08, Chakraborty09,Mendezb05a}. Still based on perturbation theory
is the $\lambda$-expansion used by Molinari in \cite{Molinari97} to deal with regular polygonal membranes. None of the
methods mentioned above has been applied to a systematic calculation of the spectrum, but rather of few selected levels. 
As a matter of fact, we are not aware of calculations in the literature for the whole spectrum of a drum, even slightly 
perturbed with respect to a solvable problem. In a different context, the study of chaos
in atomic nuclei, perturbation theory has also been used to study the statistical fluctuations of 
ground-state and binding energies~\cite{Molinari04}.

Interestingly, it seems that the analytical methods have received considerably less attention than the numerical 
methods. The development of a systematic approach to perturbation theory for membranes (quantum billiards)
of arbitrary shape is thus one the goals of the present paper. 

The paper is organized as follows: in section \ref{sec1} we describe two numerical approaches to the solution
of the Helmholz equation on arbitrary domains which are based on the use of conformal mapping (these are a new version
of the CMM of ref.~\cite{Robnik84} and the CCM of ref.~\cite{Amore08}); in section \ref{sec3} we develop two different 
analytical approaches to the solution of the problem: in the first approach, which is nonperturbative, we prove two 
theorems which provide, general expressions for the energies of the ground state or of an arbitrarily excited state; 
in the second approach, which is perturbative, we obtain a systematic expansion for the energies of an arbitrary membrane.
In section \ref{sec4} we apply the methods of sections \ref{sec1} and \ref{sec3} to several non trivial examples, which are 
helpful to illustrate the power of the techniques developed in this paper. 
Finally in section \ref{concl} we draw our conclusions.

All the numerical calculations have been carried out using Mathematica 7~\cite{Wolfram}.

\section{Numerical methods}
\label{sec1}

In this section we describe two different numerical methods: in the first part we introduce and modify the conformal 
mapping method (CMM), which was originally developed by Robnik~\cite{Robnik84}; in the second part we discuss 
the conformal collocation method (CCM), which has been developed by the author in a recent paper \cite{Amore08}.

\subsection{Conformal mapping method}
\label{CMM}

We briefly describe the conformal mapping method of Ref.~\cite{Robnik84}. The same notation of that paper
is used here. $\mathcal{D}$ is the domain of the billiard whose boundary is given by an analytic curve
in the $(u,v)$ plane. The quantum states of a particle confined in $\mathcal{D}$ are described by the hamiltonian
\beq
\hat{H} = - \frac{\hbar^2}{2m} \left[ \frac{\partial^2}{\partial u^2} + \frac{\partial^2}{\partial v^2} \right]
\label{A1}
\eeq
with Dirichlet boundary conditions on $\mathcal{D}$. In the following we will use $\hbar^2/2m = 1$. 

As explained in ref.~\cite{Robnik84} a conformal tranformation can be used to map the original region $\mathcal{D}$ 
onto the unit disk:
\beq
w = f(z)
\label{A2}
\eeq
where $w = u + i v$ and $z = x + i y$. Actually, using Riemann mapping theorem, one may choose the mapping function 
$f(z)$ so that $\mathcal{D}$ is mapped to an arbitrary simply connected region of the plane, which we call $\Omega$. 
Here we choose $\Omega$ to be a square of side $2$. Upon performing the conformal mapping the 
operator in  eq.~(\ref{A1}) transforms as
\beq
\hat{H} \rightarrow - \Sigma^{-1} \ \left[ \frac{\partial^2}{\partial x^2} + 
\frac{\partial^2}{\partial y^2} \right]  \ , 
\label{A3}
\eeq
where we have defined
\beq
\Sigma \equiv \left| \frac{df}{dz} \right|^2 \ .
\label{A4}
\eeq

We therefore obtain the eigenvalue equation
\beq
\Delta \psi(x,y) + E \Sigma(x,y) \psi(x,y) = 0 \ .
\label{A5}
\eeq

If we assume Dirichlet boundary conditions the eigenfunctions $\psi(x,y)$ may be conveniently expressed 
in the orthonormal basis of the eigenfunctions of the laplacian on the square $\Omega$ as:
\beq
\Phi_{n_x,n_y}(x,y) = \sin \left( \frac{n_x\pi}{2} (x+1)\right) \ \sin \left( \frac{n_y\pi}{2} (y+1)\right) \ ,
\label{A6}
\eeq
with $n_x, n_y = 1,2, \dots, \infty$. Von Neumann boundary conditions could be also studied by using 
the appropriate basis. We do not consider this case here.

The eigenvalues corresponding to the wavefunctions of eq.~(\ref{A6}) are 
\beq
\varepsilon_{n_x,n_y} = \frac{\pi^2}{4} \left( n_x^2+n_y^2\right) \ 
\label{A7}
\eeq
and we may write
\beq
\psi(x,y) = \sum_{n_x=1}^\infty \sum_{n_y=1}^\infty c_{n_x,n_y} \Phi_{n_x,n_y}(x,y)  \ .
\label{A8}
\eeq

By using the orthogonality of the $\Phi$ we obtain the equations
\beq
c_{n'_x,n'_y} \varepsilon_{n'_x,n'_y} + E \sum_{n_x,n_y} c_{n_x,n_y}  \Sigma_{n_x,n_y;n'_x,n'_y} = 0 
\label{A9}
\eeq
where
\beq
\Sigma_{n_x,n_y;n'_x,n'_y} \equiv \int_\Omega dx dy \ \Phi^*_{n'_x,n'_y}(x,y)  \Sigma(x,y) \Phi_{n_x,n_y}(x,y)  \ .
\label{A10}
\eeq

In order to convert eq.~(\ref{A9}) into a matrix equation we may identify a state in the box using a 
single integer index
\beq
k = n_y + N n_y \ ,
\label{A11}
\eeq
where $1 \leq k \leq N^2$ and $N^2$ is the number of states to which we have limited the sums in 
eq.~(\ref{A9})~\footnote{Of course the indices in the equation span an infinite range, however, the
numerical solution of the equation requires to work with a finite number of it.}.
We may also invert this relation to obtain
\beq
n_x = 1 + \left[ \frac{k}{N+\eta}\right] \ \  , \ \ 
n_y = k - N \left[ \frac{k}{N+\eta} \right] \ ,
\label{A12}
\eeq
where $\left[ a \right]$ is the integer part of a real number $a$ and $\eta \rightarrow 0^+$.

Using these conventions eq.~(\ref{A9}) may now be cast in the form
\beq
\left[\frac{1}{E} \ \delta_{kk'} + \frac{1}{\varepsilon_{k'}} \Sigma_{kk'} \right] c_k = 0 \ ,
\label{A13}
\eeq
with $1 \leq k \leq N^2$. 

The solution of this equation requires the calculation of a large number of integrals $\Sigma_{kk'}$, 
which is the most time consuming part of the method, as explained in Ref.~\cite{Robnik84}. 
Moreover, for large values of the indices in eq.~(\ref{A10}) the integrands are rapidly oscillating 
functions, which are more difficult to calculate numerically.

The calculation however may be done quite efficiently if the conformal map $f(z)$ is a polynomial or 
if it can be approximated with good precision by a polynomial, i.e. by the first few terms of its Taylor 
series around $z=z_0$, with $z_0 \in \mathcal{D}$. In such a case we may express $\Sigma(x,y)$ as
\beq
\Sigma(x,y) = \sum_{n=0}^\infty \sum_{m=0}^\infty \kappa_{nm} x^n y^m
\label{A14}
\eeq
and thus obtain
\beq
\Sigma_{n_x,n_y;n'_x,n'_y} = \sum_{n=0}^\infty \sum_{m=0}^\infty \kappa_{nm} \mathcal{Q}_{n_x n'_x n} \ 
\mathcal{Q}_{n_y n'_y m} \
\label{A15}
\eeq
where we have defined
\beq
\mathcal{Q}_{nmk} \equiv \int_{-1}^{+1} dx \ x^k \ \sin \left( \frac{n\pi}{2} (x+1)\right) \ \sin \left( 
\frac{m \pi}{2} (x+1)\right)  \nonumber .
\eeq

The analytical calculation of these integrals may be done quite efficiently using the recurrence relations for the 
$\mathcal{Q}_{nmk}$ given in appendix \ref{app_a}. Recurrence relations for the integrals which are relevant
for the von Neumann boundary conditions, $\mathcal{R}_{nmk}$, are also obtained.

\subsection{Conformal collocation method}
\label{sec2}

An alternative strategy for the solution of the Helmholtz equation is to use a collocation approach,
as explained in Ref.~\cite{Amore08}. We briefly review this approach.

As we have seen in the previous section the homogeneous Helmholtz equation on a domain $\mathcal{D}$ 
may be mapped to an inhogeneous Helmholtz equation on a "simpler" domain $\Omega$ (which in this 
paper we assume to be a square of side $2$). The resulting equation, eq.~(\ref{A5}), may then be
written in the equivalent form
\beq
- \frac{1}{\Sigma(x,y)} \Delta \psi(x,y) = E \psi(x,y) \ .
\label{B1}
\eeq

The discretization of this equation proceeds as follows (more detail can be found 
in  Ref.~\cite{Amore08}): first, we introduce a set of functions, the little sinc functions (LSF) 
of ref.~\cite{Amore07}, $s_k(h,N,x)$, which are defined on $x \in (-1,1)$ and obey Dirichlet bc 
at the borders (LSF corresponding to more general boundary conditions are obtained in ref.~\cite{Amore09c}) :
\beq
s_k(h,N,x) &\equiv&  \frac{1}{2 N}
\left\{ \frac{\sin \left( \left(1+\frac{1}{2N}\right) \ \frac{\pi}{h} (x-k h) \right)}{
\sin \left( \frac{\pi}{2 N h} (x-k h) \right) } \right. \nonumber \\
&-& \left. \frac{\cos\left(\left(1+\frac{1}{2N}\right) \ \frac{\pi}{h} (x+k h)\right)}{
\cos \left(\frac{\pi}{2 N h} (x+k h)\right)} \right\} \ .
\label{sincls}
\eeq
 For a given $N$ (integer)
there are $N-1$ of these functions, each peaked (with value 1) at a point $x_k = k h$ ($h=2/N$ is the
spacing of the uniform grid which these functions introduce) and vanishing at the remaining 
grid points $x_j = j h$, $j\neq k$. A function $f(x)$ obeying Dirichlet bc may be 
interpolated using the $s_k(h,N,x)$ as
\beq
f(x)  \approx \sum_{k=-N/2+1}^{N/2-1} f(x_k) s_k(h,N,x) \ .
\label{B2}
\eeq

In a similar way we may derive twice this expression to obtain
\beq
\frac{d^2f(x)}{dx^2}  &\approx& \sum_{k=-N/2+1}^{N/2-1} f(x_k) \ \frac{d^2s_k(x)}{dx^2} \nonumber \\
 &\approx& \sum_{k=-N/2+1}^{N/2-1} \sum_{j=-N/2+1}^{N/2-1} f(x_k) \ \left. \frac{d^2s_k(x)}{dx^2}\right|_{x_j} 
s_j(h,N,x) \nonumber \\
&\equiv&  \sum_{k=-N/2+1}^{N/2-1} \sum_{j=-N/2+1}^{N/2-1} f(x_k) \ c_{kj}^{(2)} \ 
s_j(h,N,x)  , 
\label{B3}
\eeq
where in the last line we have introduced the matrix $c_{kj}^{(2)} \equiv \left. \frac{d^2s_k(x)}{dx^2}\right|_{x_j}$,
which provides a representation for the second derivative operator on the grid. Notice that these 
matrix elements are known analytically. The discretization of eq.~(\ref{B1}) is now straightforward: for 
simplicity we consider the one-dimensional version of this equation and we write
\beq
-\frac{1}{\Sigma(x)} \frac{d^2}{dx^2} s_k(h,N,x) \approx - \sum_j \frac{1}{\Sigma(x_j)} \  c_{kj}^{(2)} \
s_j(h,N,x)
\label{B4}
\eeq
which provides an explicit representation of the operator $\hat{O} \equiv -\frac{1}{\Sigma(x)} \frac{d^2}{dx^2}$ on the
uniform grid.

If we want to generalize these results to two dimensions, we may consider the set of functions
which are obtained from the direct product of the set of functions in the $x$ and $y$ directions, 
i.e. $s_k(h,N,x) s_{k'}(h,N,x)$. Notice that we are assuming that the number of elements in the two directions
is the same although different values could be considered if needed. Clearly the representation
of an operator on the grid in terms of these functions will depend on four indices (the two integer indices
corresponding to the initial point on the grid and the two integers indices corresponding to the 
final point on the grid). However we may identify a specific point on the grid with a single integer
\beq
K = k' + \frac{N}{2} + (N-1) \left(k + \frac{N}{2}-1\right) \ .
\eeq
where $1 \leq K \leq (N-1)^2$. This relation is easily inverted to give:
\beq
\label{B5a}
k  &=& 1 - \frac{N}{2} + \left[ \frac{K}{N-1+\eta}\right] \\
k' &=& K - \frac{N}{2} - (N-1) \left[ \frac{K}{N-1+\eta}\right] \ ,
\label{B5b}
\eeq 
where $\left[a\right]$ is the integer part of $a$ and $\eta \rightarrow 0^+$.

Using these results one may easily generalize eqn.~(\ref{B4}) to two dimensions as
\beq
&-& \frac{1}{\Sigma(x,y)} \Delta  s_k(h,N,x) s_{k'}(h,N,y) = - \sum_{jj'} \frac{1}{\Sigma(x_j,y_{j'})} \nonumber \\
&\times& \left[ c_{kj}^{(2)} \delta_{k'j'} + \delta_{kj} c_{k'j'}^{(2)} \right]  s_j(h,N,x) s_{j'}(h,N,y) 
\eeq
from which we easily obtain the representation of $\hat{O} \equiv - \frac{1}{\Sigma(x,y)} \Delta $ on the 
grid~\footnote{Eqns. (\ref{B5a}) and (\ref{B5b}) and the corresponding equations for the indices $j$ and $j'$
allow to express everything in terms of $K$ and $K'$, since $k=k(K)$, $k'=k'(K)$ and $j=j(K')$ and $j'=j'(K')$.}:
\beq
O_{KK'} = -  \frac{1}{\Sigma(x_j,y_{j'})}  \left[ c_{kj}^{(2)} \delta_{k'j'} + \delta_{kj} c_{k'j'}^{(2)} \right]  \ .
\eeq

We underline some useful features of this expression:
\begin{itemize}
\item the matrix corresponding to $\left[ c_{kj}^{(2)} \delta_{k'j'} + \delta_{kj} c_{k'j'}^{(2)} \right]$ 
is the representation of the 2D laplacian operator on the uniform grid in a square of side $L=2$; 
this matrix is "universal", i.e. not specific to the particular problem one is solving, and sparse.
\item the matrix corresponding to $\frac{1}{\Sigma(x_j,y_{j'})}$ is specific to the domain $\mathcal{D}$
since $\Sigma$ is related to the conformal map; however this matrix is diagonal and therefore only $N-1$ 
elements need to be calculated; 

\item the matrix $O$ is a {\sl nonsymmetrical} real square matrix whose eigenvalues are real.

\item no integrals need to be calculated, since $\Sigma$ is simply evaluated on the grid points. In 
this case one {\sl does not need} to approximate the map with a polynomial in $z$.
\end{itemize}

Because of the features above, the optimal computational strategy consists of calculating the 
matrices for the discretized laplacian on $\Omega$ for grids of different sizes. 
Typically, this is the most time consuming task but it only needs to be done once, since 
these matrices are common to all problems and therefore they can be stored and used when 
needed without having to recalculate them. 
When a particular conformal map is chosen, one only needs to calculate the $N-1$ diagonal 
elements of $\Sigma$, which has a much more limited computational cost. 
The matrix representation for $\hat{O}$ is thus obtained quite effectively and 
selected eigenvalues (eigenvectors) may be calculated rapidly. We will discuss some applications
of this approach in Section \ref{sec4}.

\section{Analytical methods}
\label{sec3}

In this section we derive useful analytical approximations for the
energies of a particle confined in a two dimensional region. In the first
part we obtain a systematic approximation to the ground state energy and 
wave function (or more in general to the lowest state in each symmetry class of 
the problem): this expansion is really non-perturbative and it is proved
to converge to the exact values; in the second part we obtain the analytical 
expression for the spectrum of an arbitrary two dimensional region, 
worked out in perturbation theory up to third order.

\subsection{Nonperturbative formulas}
\label{subsec1}

We will now derive a general formula for the ground state of a drum obeying
Helmholtz equation (\ref{A5}) or the equivalent equation (\ref{B1}).

As noticed before the conformal map has allowed to convert the original homogeneous 
Helmholtz equation, defined on a domain $\mathcal{D}$, into an inhomogenous 
Helmholtz equation, defined on a "simpler" domain $\Omega$, which in the present 
case we will assume to be a square of side $2$. 

We make few observations regarding the composite operator 
$\hat{O} = - \frac{1}{\Sigma(x,y)} \Delta$, which will help us to devise a suitable 
approach to finding approximations to the states of eq.(\ref{B1}): 
\begin{itemize}
\item the operator $\hat{O}$ is non-hermitian, $\hat{O}^\dagger \neq \hat{O}$. This
      confirms the observation made in the previous section that the matrix representing 
      the operator on the uniform grid is real and non-symmetrical. 
\item the operator $\hat{O}$ may be cast in a manifestly hermitian form by considering its symmetrized 
      form 
\beq
      \hat{O}^{(sym)} = \frac{1}{\sqrt{\Sigma}} (-\Delta) \frac{1}{\sqrt{\Sigma}} \ .
\eeq
      In the framework of the Conformal Collocation Method (CCM) of the previous section, the matrix 
      representing this operator on a uniform grid is symmetrical and  hermitian. 
      Moreover its eigenvalues coincide with those of the unsymmetrized matrix. From now on we will 
      always work with $\hat{O}^{(sym)}$ and therefore we will omit the superscript.
      It is interesting to notice that the form that we have found for $\hat{O}^{(sym)}$ has been
      proposed by Zhu and Kroemer in \cite{Zhu83}, discussing the problem of the connection rules
      for effective mass wave functions across abrupt heterojunctions. A discussion of 
      kinetic operators containing more general position dependent effective mass terms is 
      discussed in \cite{vonRoos83}. 

\item $\Omega$ is a two-dimensional domain on which the spectrum (energies and wave functions) 
      of the negative laplacian operator is known exactly, i.e. a square (rectangle) or a circle.

\item the eigenvalues of $\hat{O}$ are definite positive if we assume Dirichlet boundary conditions: 
      the inverse operator $\hat{O}$ exists. For von Neumann boundary conditions a state with zero 
      energy is present and one needs to define the invertible operator 
\beq
\hat{O}_\eta =  \frac{1}{\sqrt{\Sigma}} (-\Delta + \eta) \frac{1}{\sqrt{\Sigma}} \ .
\eeq      
      where $\eta \rightarrow 0^+$ is a positive infinitesimal parameter.

\item the spectrum of $\hat{O}$ is not bounded from above, i.e. there is an infinite number of 
      states, of arbitrarily high energy;

\item the spectrum of the inverse operator $\hat{O}^{-1}$ is {\sl bounded from above}, the
      largest eigenvalue being the reciprocal of the lowest eigenvalue of $\hat{O}$ (of course 
      there is still an infinite number of states which become denser and denser as the eigenvalues
      approach zero). The spectrum of $\hat{O}_\eta^{-1}$ corresponding to von Neumann boundary conditions
      is also   {\sl bounded from above}, although its largest eigenvalue tends to infinity as
      $\eta$ tends to zero.
\end{itemize}

To simplify the notation in the following we will write $\hat{O}_\eta$ as simply $\hat{O}$, 
implicitly assuming that an infinitesimal positive shift $\eta$ is used if von Neumann bc are used.

We are now in position of stating the following Theorem:
\begin{theorem}
Let $|\Psi_0\rangle$ be the exact ground state of the operator $\hat{O}$ and
$|\Psi\rangle$ be an arbitrary state with nonzero overlap with $|\Psi_0\rangle$. 
Then the lowest mode of the Helmholtz equation (\ref{B1}) is given by
\beq
| \Psi_0 \rangle = \lim_{n\rightarrow \infty} \left[\hat{O}^{-1}\right]^n |\Psi\rangle \ .
\label{C2}
\eeq
where $\hat{O}^{-1} = \sqrt{\Sigma} (-\Delta+\eta)^{-1} \sqrt{\Sigma}$ and
$\Sigma \equiv \left| \frac{df}{dz} \right|^2$ ($f(z)$ is the conformal map which 
maps $\mathcal{D}$ into $\Omega$). More in general, if $| \Psi\rangle$ is orthogonal 
to the first $n$ states of eq.(\ref{B1}) then the resulting state converges to the $n+1$
excited state.
\end{theorem}

The proof of this theorem is easy: for simplicity we limit to Dirichlet bc and therefore set 
$\eta = 0$ (the case of von Neumann bc is obtained straightforwardly, working with a nonzero 
$\eta$).

Given an arbitrary state $|\Psi\rangle$, we may decompose
it in terms of the eigenstates of $\hat{O}$, $|\Psi_n\rangle$, and write 
$|\Psi\rangle = \sum_{n=0}^\infty a_n |\Psi_n\rangle$.

By applying the inverse operator to this state we obtain
\beq
\hat{O}^{-1} |\Psi\rangle =  \sum_{n=0}^\infty \frac{a_n}{E_n} |\Psi_n\rangle
\label{C3}
\eeq
where $E_n$ are the eigenvalues of $\hat{O}$. These eigenvalues are positive definite, $E_n > 0$,
for all values of $n$. For this reason the component of $|\Psi_0\rangle$ in this
new state has been amplified with respect to all the other components  $|\Psi_{n\geq 1}\rangle$. 
Therefore repeated applications of the inverse operator  will further suppress these components 
and the resulting state will converge to the true ground state of $\hat{O}$. This concludes
our demonstration.

What we have described so far is actually an implementation of the well known {\sl Power Method}~\cite{OLeary79} 
for an infinite dimensional matrix, with positive definite eigenvalues which are bounded from above.

There are several aspects which make the present implementation of the Power Method particularly appealing:
\begin{itemize}
\item the convergence rate of the algorithm depends on the ratio of $E_1/E_0$, becoming slower when $E_0$ and $E_1$ 
      are not well separated. An upper bound to this ratio is provided by the Payne-Polya-Weinberger conjecture~\cite{PPW1,PPW2}
\beq
\frac{E_1}{E_0} \leq \left.\frac{E_1}{E_0} \right|_{disk} = \left(\frac{j_{1,1}}{j_{0,1}}\right)^2 \approx 2.539
\label{C4}
\eeq
where $j_{0,1}$ and $j_{1,1}$ are the first positive zeroes of the Bessel functions $J_0(x)$ and $J_1(x)$.
This conjecture has been proved in ref.~\cite{Benguria91}. On the other hand, it is easy to convince oneself that
the ratio will get close to $1$ for elongated domains $\mathcal{D}$, where the longitudinal dimension is much 
larger then the transverse one. In these cases one may combine the method with a suitable dilatation 
in order to improve the convergence.

\item the ground state of the negative laplacian on $\Omega$ provides an initial guess
of good quality (the typical implementation of the Power Method uses a random initial guess);

\item in the case of von Neumann bc the initial guess should be chosen orthogonal
to the zero energy state, which is trivial. 

\item we may reasonably expect that the lowest modes of the negative laplacian on $\Omega$, 
which we call $|n\rangle$, dominate in $|\Psi_0\rangle$;  therefore we may leave the coefficients 
of the expansion of the initial  guess in terms of the $|n\rangle$ unspecified and use the variational 
principle to obtain them. In other words we may look for the coefficients which
minimize the expectation value of $\hat{O}$ in the state generated with the Power Method.
\end{itemize}

We now explore explicitly the application of the variational principle to first order. For simplicity
we assume $\Omega$ to be a square.  Let us choose an arbitrary state $|\chi\rangle$ which can be decomposed 
in terms of the eigenstates of the Laplacian on a box as
\beq
| \chi \rangle = \sum_k c_k^{(0)} |k\rangle .
\label{C5}
\eeq

Notice that the state $|k\rangle = |k_x,k_y\rangle$ is the direct product of the states
corresponding to each direction. Similarly we call $\epsilon_k$ the eigenvalue of the negative
Laplacian on $\Omega$ for  this state, $\epsilon_k \equiv  \frac{\pi^2}{4} \left( k_x^2+k_y^2\right)$.

Working to first order we obtain an explicit approximate expression for the ground state 
\beq
| \Psi_0^{(1)} \rangle &=& \left[\hat{O}^{-1}\right] |\chi \rangle 
= \left[ \Sigma^{1/2} \left(-\Delta\right)^{-1} \Sigma^{1/2} \right] | \chi\rangle  \nonumber \\
&=& \sum_{k,l,m=0}^\infty  \frac{c_m^{(0)}}{\epsilon_l} |k \rangle \langle k | \Sigma^{1/2} | l \rangle 
\langle l | \Sigma^{1/2} | m \rangle \ .
\eeq
We may cast this result in the form
\beq
| \Psi_0^{(1)} \rangle &=& \sum_{k} c_k^{(1)} | k\rangle
\label{C6}
\eeq
where
\beq
c_k^{(1)} &\equiv& \sum_{l,m=0}^\infty  \frac{c_m^{(0)}}{\epsilon_l} \langle k | \Sigma^{1/2} | l \rangle 
\langle l | \Sigma^{1/2} | m \rangle \ .
\label{C7}
\eeq

Using this result we may obtain the first order approximation to the ground state 
energy~\footnote{The reader may notice that a variational bound on the ground state energy could also have been obtained 
calculating the expectation value of $\hat{O}$ in  $|\chi\rangle$: 
\beq
\frac{\langle \chi | \hat{O} |\chi \rangle}{\langle \chi|\chi \rangle} &=& 
\frac{\sum_{k,l,m=0}^\infty  c_k^{(0)} c_m^{(0)} \epsilon_l  \langle m | \Sigma^{-1/2} | l \rangle  
\langle l | \Sigma^{-1/2} | k \rangle}{\sum_{k=0}^\infty \left. c_k^{(0)}\right.^2 } \ , \nonumber
\eeq
and then minimizing this expression with respect to the $c_k^{(0)}$.
This expression, however, has two unpleaseant features: first, it contains matrix elements of the operator $\Sigma^{-1/2}$,
instead of $\Sigma^{1/2}$; second, and most important, it contains the energies of internal states in the numerator, 
instead that in the denominator. Since in any practical application of these formulas a cutoff is used for 
the internal sums, this expression will be more sensitive to this cut and therefore less precise.}:
\beq
E_0^{(1)} &=& \frac{\langle \Psi_0^{(1)} | \hat{O} | \Psi_0^{(1)}\rangle}{\langle \Psi_0^{(1)} | 
\Psi_0^{(1)}\rangle} \ ,
\label{C8}
\eeq
where
\beq
\langle \Psi_0^{(1)} | \hat{O} | \Psi_0^{(1)}\rangle &=& 
\sum_{k,l,m=0}^\infty  \frac{c_k^{(0)} c_m^{(0)}}{\epsilon_l } \langle m | \Sigma^{1/2} | l \rangle  
\langle l | \Sigma^{1/2} | k \rangle \\
\langle \Psi_0^{(1)} | \Psi_0^{(1)}\rangle &=& 
\sum_{l,l',m,m'=0}^\infty  \frac{c_m^{(0)} c_{m'}^{(0)}}{\epsilon_l \epsilon_{l'} } \langle m' | \Sigma^{1/2} | l' \rangle
\langle l' | \Sigma | l \rangle  \nonumber \\
&\times& \langle l | \Sigma^{1/2} | m \rangle \ .
\label{C9}
\eeq

$E_0^{(1)}$ given above provides an upper bound to the true energy $E_0$:
\beq
E_0^{(1)} \geq E_0 .
\label{C10}
\eeq

The coefficients $c_n^{(0)}$ may therefore be chosen so that the expectation value of $\hat{O}$ is minimized.
Therefore using a suitable number of coefficients in principle one may obtain arbitrarily accurate estimates 
for $E_0$. 

If we are willing to trade the accuracy for the simplicity, we may set $c_0^{(0)} = 1$ and $c^{(0)}_{n\geq 1}=0$: 
in this case we obtain the approximation
\beq
E_0^{(1)} &=& \frac{\sum_{l=0}^\infty  \frac{1}{\epsilon_l } \langle 0 | \Sigma^{1/2} | l \rangle  
\langle l | \Sigma^{1/2} | 0 \rangle }{\sum_{l,l'=0}^\infty  \frac{1}{\epsilon_l \epsilon_{l'} } 
\langle 0 | \Sigma^{1/2} | l' \rangle \langle l' | \Sigma | l \rangle \langle l | \Sigma^{1/2} | 0 \rangle} 
\label{C11}
\eeq
which reduces to a quite simple formula when the internal sums are limited to the lowest state
\beq
E_0^{(1)} \approx \frac{\epsilon_0}{\langle 0| \Sigma | 0 \rangle } \ .
\label{C12}
\eeq

In Section \ref{sec4} we will consider few applications of the formulas given above.

The generalization of these results to higher orders is trivial after observing that
\beq
| \Psi_0^{(n)} \rangle &=& \left[\hat{O}^{-1}\right] |\Psi_0^{(n-1)} \rangle = \sum_{k} c_k^{(n)} | k\rangle
\label{C13}
\eeq
where
\beq
c_k^{(n)} &\equiv& \sum_{l,m=0}^\infty  \frac{c_m^{(n-1)}}{\epsilon_l} \langle k | \Sigma^{1/2} | l \rangle 
\langle l | \Sigma^{1/2} | m \rangle \ .
\label{C14}
\eeq

This equation provides a useful recurrence relation for the coefficients $c_k$.

For example we have
\beq
c_k^{(2)} &\equiv& \sum_{l,m=0}^\infty  \frac{c_m^{(1)}}{\epsilon_l} \langle k | \Sigma^{1/2} | l \rangle 
\langle l | \Sigma^{1/2} | m \rangle \nonumber \\
&=& \sum_{l,m=0}^\infty  \frac{c_m^{(0)}}{\epsilon_l\epsilon_{l'}} \langle k | \Sigma^{1/2} | l \rangle 
\langle l | \Sigma | l' \rangle \langle l' | \Sigma^{1/2} | m \rangle \ ,
\label{C15}
\eeq
where the completeness of states $\sum |k\rangle \langle k |=1$ has been used to obtain the final expression.


The results that we have just obtained for the ground state may also be generalized for the excited states.
As a matter of fact we may state the following Theorem:
\begin{theorem}
Let $|\Psi_n^{(d)}\rangle$ be an eigenstate of the operator $\hat{O}$ with degeneracy $d$ and 
with an eigenvalue $E_n$. Let $|\Psi\rangle$ be an arbitrary state with nonzero overlap with 
at least one of the $|\Psi_n^{(d)}\rangle$ and let $\Lambda$ be a real parameter for which 
$|E_n-\Lambda|\ll 1$. Then the state
\beq
| \Phi \rangle = \lim_{n\rightarrow \infty} \left[\left(\hat{O}-\Lambda\right)^{-2}\right]^n |\Psi\rangle \ .
\label{theo2}
\eeq
is a linear combination only of the $d$ degenerate states $|\Psi_n^{(d)}\rangle$.
\end{theorem}

The proof of this theorem is analogous to the proof that we have given earlier for the ground state. 
We just need to notice that the eigenvalues of the operator $\left(\hat{O}-\Lambda\right)^{-2}$ 
are positive definite and bounded from above. Because of the condition $|E_n-\Lambda|\ll 1$ 
the repeated application of this operator to $|\Psi\rangle$ amplifies the components corresponding 
to the eigenvalue $E_n$ since $1/(E_n-\Lambda)^2$ is the largest eigenvalue of 
$\left(\hat{O}-\Lambda\right)^{-2}$. When the limit is taken only the components corresponding
to this degenerate eigenvalue survive, which proves the theorem. By choosing other 
$d-1$ linearly indipendents ansatz and repeating the procedure, we obtain $d$ linearly independent combinations
of the degenerate states, which we can use to calculate the matrix elements of $\hat{O}$. The diagonalization
of this matrix provides the $d$ eigenstates of interest.
Notice that this strategy for calculating arbitrary eigenvalues and eigenvectors of large matrices is
discussed in refs.~\cite{Nash79,WZ94,GMP95}.

\subsection{Perturbation theory}
\label{subsec2}

While the previous approach based on the power method provides a systematic approximation to the 
ground state of a drum of arbitrary shape, an alternative approach consists of applying perturbation 
theory to obtain the corrections to the energies and wave functions. 

Once again our starting point is the symmetrized operator
\beq
\hat{O} = \frac{1}{\sqrt{\Sigma}} (-\Delta) \frac{1}{\sqrt{\Sigma}} \ ,
\label{pt1}
\eeq
which reduces to the negative laplacian for $\Sigma =1$, corresponding to a square of side $L=2$. 
If we consider small deformations of this square and write 
\beq
\Sigma \rightarrow \Sigma_\eta = 1 + \eta \sigma
\label{pt2}
\eeq
where $\sigma \equiv \Sigma-1$ and $\eta$ is a power counting parameter (which at the end of the calculation 
is set to $1$), which is used to keep track of the different orders in powers of $\sigma$. 

Our operator may now be expanded in powers of $\eta$ as
\beq
\hat{O} &\approx& \hat{O}_0 + \eta \hat{O}_1 + \eta^2 \hat{O}_2 + \eta^3 \hat{O}_3 + \dots \ ,
\label{pt3}
\eeq
where the explicit form of the  $\hat{O}_i$ may be worked out rather easily:
\beq
\label{pt4}\hat{O}_0 &=& -\Delta \\
\label{pt5}\hat{O}_1 &=& - \frac{1}{2} \left[ \sigma (-\Delta) + (-\Delta) \sigma\right] \\
\label{pt6}\hat{O}_2 &=& \frac{1}{8} \left[ 2 \sigma (-\Delta) \sigma + 3 \sigma^2 (-\Delta) + 
3 (-\Delta) \sigma^2 \right]  \\
\hat{O}_3 &=&  - \frac{3}{16} \left[  \sigma^2 (-\Delta) \sigma +  \sigma (-\Delta) \sigma^2 \right]  \nonumber \\
\label{pt7}&-& \frac{5}{16} \left[  \sigma^3 (-\Delta)  +   (-\Delta) \sigma^3 \right]  
\eeq

Within perturbation theory we may now calculate the corrections to the eigenvalues
and eigenfunctions of our operator $\hat{O}$, keeping in mind that it contains 
arbitrary  powers of $\eta$. The standard scheme of Rayleigh-Schr\"odinger perturbation
theory (RSPT) may be easily adapted to this problem and the contributions up to third order read:
\beq
\label{pt8}E_n^{(0)} &=& \epsilon_n \\
\label{pt9}E_n^{(1)} &=& \langle n | \hat{O}_1| n \rangle \\
\label{pt10}E_n^{(2)} &=& \langle n | \hat{O}_2| n \rangle + 
\sum_{k\neq n} \frac{|\langle n | \hat{O}_1 | k \rangle|^2}{\epsilon_n -\epsilon_k} \\
E_n^{(3)} &=& \langle n | \hat{O}_3| n \rangle + 2
\sum_{k\neq n} \frac{\langle n | \hat{O}_2 | k \rangle \langle k | \hat{O}_1 | n \rangle}{\epsilon_n -\epsilon_k} 
\nonumber \\
&+& \sum_{k\neq n} \sum_{m \neq n} 
\frac{\langle n | \hat{O}_1 | m \rangle \langle m | \hat{O}_1 | k \rangle \langle k | 
\hat{O}_1 | n \rangle}{(\epsilon_n -\epsilon_k) (\epsilon_n -\epsilon_m)} \nonumber \\
\label{pt11}&-& \langle n | \hat{O}_1 | n \rangle 
\sum_{k\neq n} \frac{ \langle n | \hat{O}_1 | k \rangle^2}{(\epsilon_n -\epsilon_k)^2} \ .
\eeq

Notice that the expressions corresponding to orders $\eta^2$ and $\eta^3$ receive contributions
from distinct operators $\hat{O}_i$. 

We have calculated the perturbative corrections to the energy up to third order for the non-degenerate
part of the spectrum:
\beq
\label{pt12}E_n^{(1)} &=& - \epsilon_n \langle n | \sigma | n \rangle  \\
\label{pt13}E_n^{(2)} &=& \epsilon_n \langle n | \sigma | n \rangle^2 
+ \epsilon_n^2 \sum_{k \neq n} \frac{\langle n | \sigma | k \rangle^2 }{\epsilon_n-\epsilon_k} \\
E_n^{(3)} &=& - \epsilon_n \langle n | \sigma | n \rangle^3 + \epsilon_n^3 \langle n | \sigma | n \rangle
\sum_{k \neq n}  \frac{\langle n | \sigma | k \rangle^2}{\omega_{nk}^2} \nonumber \\
&-& 3 \epsilon_n^2  \langle n | \sigma | n \rangle \sum_{k \neq n}  
\frac{\langle n | \sigma | k \rangle^2}{\omega_{nk}} \nonumber \\
\label{pt14}&-& \epsilon_n^3 \sum_{k\neq n} \sum_{m \neq n} 
\frac{\langle n | \sigma | k \rangle \langle k | \sigma | m \rangle 
\langle m | \sigma | n \rangle }{\omega_{nk}\omega_{nm}}  \ .
\eeq
Details of the derivation of these coefficients are given in the Appendix \ref{app_b}.

The expressions above seem to suggest the presence of a term $(-1)^k \langle n | \sigma | n \rangle^k$
in the perturbative coefficient of order $k$: assuming that this is true, one recovers a geometric series
which produces a term $\epsilon_n/(1+ \langle n | \sigma | n \rangle)$,
which, for $n=0$, is precisely the term obtained for the ground state energy working to lowest order, eq.~(\ref{C12}).

We may now discuss the extension of these results for degenerate states. Suppose that a state 
is $d$ times degenerate: in this case the perturbative expansion needs then  to be carried out 
using  the eigenstates of the $d \times d$ matrix $\overline{\sigma}_d$ of elements 
$\langle n_i | \sigma | n_j\rangle$, where $|n_i\rangle$ ($i=1,\dots, d$) is one of the degenerate states. 
With this modification the sums over internal states automatically exclude the degenerate
states, given that the interaction does not mix them, and the perturbation expansion is well defined. 
Notice that if the matrix $\overline{\sigma}_d$ is diagonal and proportional to the identity matrix, the degeneracy of the 
levels  will not be lifted by perturbation theory. We will discuss some examples of this in the Section \ref{sec4} 
considering the spectrum of a circular membrane.

To illustrate how this works in practice we may consider the simplest case, in which the states are twice degenerate.
Let us call $|n_1\rangle$ and $|n_2\rangle$ these states and $\epsilon_1$ and $\epsilon_2$ their corresponding 
energies. If we work with perturbations of the square, the states of double degeneracy (which are also the
most common states) are states with quantum numbers $(n_x,n_y)$ and $(n_y,n_x)$.

The eigenvalues of the matrix $\overline{\sigma}_{d=2}$ are:
\beq
\tilde{\varepsilon}_{1,2} &=& \frac{1}{2} \left[ \langle n_1 | \sigma| n_1\rangle +
\langle n_2 | \sigma| n_2\rangle \pm \xi_{12} \right]
\eeq
where
\beq
\xi_{12}\equiv \sqrt{\left(\langle n_1 | \sigma| n_1\rangle -\langle n_2 | \sigma| n_2\rangle\right)^2 + 4 
\langle n_1 | \sigma| n_2\rangle^2} \ .
\eeq

We clearly see that if $\langle n_1 | \sigma| n_2\rangle =0$, then the $\tilde{\varepsilon}_{1,2}$ are degenerate
(as anticipated). The eigenvectors of $\overline{\sigma}_{d=2}$ are:

Its eigenvectors are:
\beq
|\tilde{n}_{1,2}\rangle  &=& \beta_1^{(1,2)} | n_1\rangle + \beta_2^{(1,2)} | n_2\rangle  \ ,
\eeq
where
\beq
\beta_{1}^{(1)} &=& -\frac{1}{\sqrt{2}} \ \sqrt{1 - \frac{\langle n_1 | \sigma| n_1\rangle -
\langle n_2 | \sigma| n_2\rangle}{\xi_{12}}} \\
\beta_{2}^{(1)} &=& \frac{1}{\sqrt{2}} \ \sqrt{1 + \frac{\langle n_1 | \sigma| n_1\rangle -
\langle n_2 | \sigma| n_2\rangle}{\xi_{12}}} \\
\beta_{1}^{(2)} &=& \frac{1}{\sqrt{2}} \ \sqrt{1 + \frac{\langle n_1 | \sigma| n_1\rangle -
\langle n_2 | \sigma| n_2\rangle}{\xi_{12}}} \\
\beta_{2}^{(2)} &=& -\frac{1}{\sqrt{2}} \ \sqrt{1 - \frac{\langle n_1 | \sigma| n_1\rangle -
\langle n_2 | \sigma| n_2\rangle}{\xi_{12}}}  \ .
\eeq

We may now explicitly write down the different orders in PT. For example to first order we have
\beq
E_n^{(1)} &=& - \epsilon_{n} \langle \tilde{n}_{1,2} | \sigma | \tilde{n}_{1,2} \rangle  \nonumber  \\
&=&  - \frac{\epsilon_n}{2}  \left[ \langle n_1 | \sigma| n_1\rangle +\langle n_2 | \sigma| n_2\rangle 
\pm \xi_{12} \right]  \ .
\eeq

To second order we have:
\beq
E_n^{(2)} &=& \epsilon_n \langle \tilde{n}_{1,2} | \sigma | \tilde{n}_{1,2} \rangle^2
+ \epsilon_n^2 \sum_{k \neq n} \frac{\langle \tilde{n}_{1,2} | \sigma | k \rangle^2 }{\epsilon_n-\epsilon_k} \nonumber \\
&=& \frac{\epsilon_n}{4}  \left[ \langle n_1 | \sigma| n_1\rangle + \langle n_2 | \sigma| n_2 \rangle \pm 
\xi_{12} \right]^2 
\nonumber \\ 
&+&  \epsilon_n^2  \sum_{k \neq n} \frac{\left(\beta_1^{(1,2)} \langle n_{1} | \sigma | k \rangle+
\beta_2^{(1,2)} \langle n_{2} | \sigma | k \rangle\right)^2 }{\epsilon_n-\epsilon_k} 
\eeq

We do not write explicitly the formula of the third order which is obtained in a similar fashion.

\subsection{Perturbation theory again}
\label{subsec3}

Let us now try to obtain perturbative expressions for the energies using a different approach, which
uses the theorems of eq.~(\ref{C2}) and (\ref{theo2}). For simplicity we limit our considerations only
to the ground state.

Using eq.~(\ref{C2}) we write 
\beq
E_0 = \lim_{n \rightarrow \infty} \frac{\langle \Psi | \left[O^{-1} \right]^{n-1} | \Psi\rangle}{\langle \Psi | 
\left[O^{-1} \right]^{n} |\Psi\rangle} \ ,
\label{newpert1}
\eeq
where
\beq
\hat{O}^{-1} = \sqrt{1+\eta \sigma} \ \left(-\hat{\Delta}^{-1}\right) \ \sqrt{1+\eta \sigma}  \ .
\eeq
and $|\Psi\rangle$ is an arbitrary state with non-zero overlap with the exact ground state.

The perturbative series for $E_0$ is of the form
\beq
E_0^{(PT)} = \sum_{k=0}^\infty E_0^{(k)} \eta^k 
\eeq
and therefore
\beq
E_0^{(k)} = \frac{1}{k!} \left. \frac{\partial^k E_0^{(PT)}}{\partial \eta^k} \right|_{\eta=0} \ .
\eeq

In taking the derivatives of the expression (\ref{newpert1}) we need to consider that the state $|\Psi\rangle$
is {\sl arbitrary} and therefore it may be chosen to be independent of $\eta$. A suitable choice is to 
pick $|\Psi\rangle$ the ground state of the unperturbed hamiltonian, $-\Delta$.

Let us consider, for example, the first order term:
\begin{widetext}
\beq
E_0^{(1)} &=& \left. \frac{\partial E_0^{(PT)}}{\partial \eta} \right|_{\eta=0}  
= \left. \lim_{n \rightarrow \infty} \left[\frac{\langle \Psi | \frac{\partial}{\partial\eta} \left[O^{-1} \right]^{n-1} 
| \Psi\rangle}{\langle \Psi | \left[O^{-1} \right]^{n} |\Psi\rangle} - 
\frac{\langle \Psi | \left[O^{-1} \right]^{n-1} | \Psi\rangle
\langle \Psi | \frac{\partial}{\partial\eta} \left[O^{-1} \right]^{n} 
| \Psi\rangle}{\langle \Psi | \left[O^{-1} \right]^{n} |\Psi\rangle^2} \right] \right|_{\eta=0}  \ .
\eeq
\end{widetext}

In order to evaluate this expression we only need to consider the terms containing the derivatives, since
the remaining terms may already be evaluated at $\eta=0$. For instance
\beq
\left. \langle \Psi | \left[O^{-1} \right]^{n} |\Psi\rangle\right|_{\eta=0} = \left[E_0^{(0)}\right]^{-n} \ .
\eeq

Let us now consider $\hat{O}^{-n}$ us to order $\eta$:
\beq
\hat{O}^{-n} &\approx& (-\Delta)^{-n} + \eta \left[ \frac{\sigma}{2} (-\Delta)^{-n} + 
(-\Delta)^{-1} \sigma (-\Delta)^{-n+1} \right. \nonumber \\ 
&+& \left. \dots + (-\Delta)^{-n+1} \sigma (-\Delta)^{-1} + (-\Delta)^{-n} \frac{\sigma}{2}
\right] \nonumber
\eeq

Therefore
\beq
\left.\langle \Psi | \frac{\partial}{\partial\eta} \left[O^{-1} \right]^{n} | \Psi\rangle \right|_{\eta=0} =
(n+1) \langle \Psi | \sigma |\Psi\rangle  \left[E_0^{(0)}\right]^{-n} \ .
\eeq

We thus obtain:
\beq
E_0^{(1)} &=& \lim_{n \rightarrow \infty} \left[  n E_0 \langle \Psi | \sigma |\Psi\rangle - (n+1) E_0 
\langle \Psi | \sigma |\Psi\rangle \right] \nonumber \\
&=& -  E_0 \langle \Psi | \sigma |\Psi\rangle \ ,
\eeq
which agrees with the formula obtained earlier.

\section{Applications}
\label{sec4}

In this section we consider several applications of the methods described in the previous sections and compare
the results obtained with those in the literature. 

\subsection{Unit circle}

\begin{figure}
\begin{center}
\bigskip\bigskip\bigskip
\includegraphics[width=5cm]{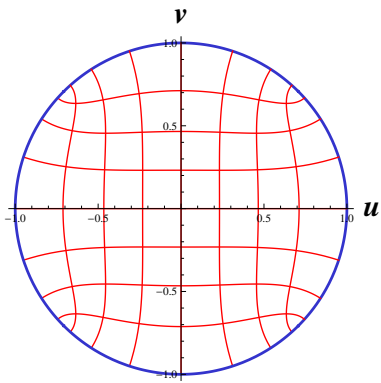}
\caption{(color online) Unit circle obtained from the conformal map of the square of side $2$.}
\label{Fig_0}
\end{center}
\end{figure}

The exact wave functions and energies of a circular billiard are known analytically and are expressed 
in terms of the Bessel functions and its zeroes. This example therefore provides an ideal test both 
for our numerical methods and for our analytical approximations (clearly $\Omega$ in this case is 
a square). 
Fig.~\ref{Fig_0} displays the unit circle which is obtained from a square of side $2$ using a conformal map. 
The internal lines in the circle correspond to the mapping of a uniform grid on the square. 
Here the map has been approximated with the Taylor expansion to order $z^{37}$. The effect of this approximation 
is well under control, since the coefficients decay quickly: for instance, the coefficient of $z^{37}$ is  
$1.6 \times 10^{-11}$. We may roughly expect that the smallest errors that we can obtain in our calculation stopping 
to this order are of this order of magnitude. Notice that using higher orders in the Taylor expansion makes sense only if 
one can handle the matrices of the size needed to achieve the extra precision made available by the Taylor expansion.

We have applied both methods described in this paper to this problem, working to different orders. The results
obtained for the first three levels are reported in Tables \ref{table-1} (CMM) and \ref{table-2} (CCM). In the case
of CCM $N^2$ is the number of elements of the basis of the square which have been used; in the case of the CCM
$N^2$ is the number of internal grid points used for the discretization. The reader may notice that 
the results obtained with the CMM for these states are more precise than those of the CCM for the same states: 
moreover the lowest error reached with the CMM is of the order of $10^{-11}$.

A second feature, which is common to both methods, is that the sequence of values obtained for a given
eigenvalue changing $N$ is monotonously decreasing, as one would expect from the variational 
principle. This implies that the extrapolation of values corresponding to different $N$ may be
used to obtain more accurate results. Notice that this is not generally true for finite element methods, which
do not describe exactly the border of the membrane.

\begin{figure}
\begin{center}
\bigskip\bigskip\bigskip
\includegraphics[width=7cm]{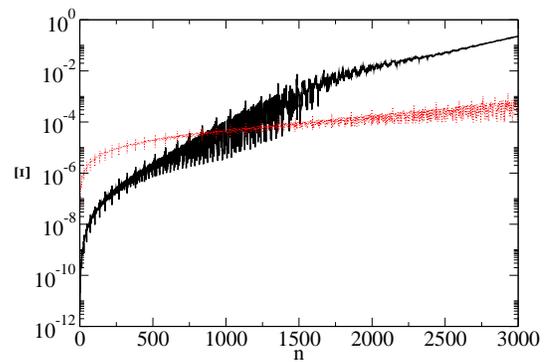}
\caption{(color online) Error over the first $3000$ energy levels of the circular membrane of radius $r=1$ 
obtained using CMM with $N=60$ (thick line) and the CCM with $N=100$ (thin line).}
\label{Fig_0b}
\end{center}
\end{figure}

In Fig.~\ref{Fig_0b} we have plotted the error $\Xi \equiv \frac{E_n^{approx}}{E_n^{exact}}-1$  
for the first $3000$ energy levels of the circle using either the CMM with $N=60$ (thick black line) or 
the CCM with $N=100$ (thin red line). We need to make two observations at this point: first, that the problem
is symmetric with respect to both axes and therefore one could use this property to reduce the dimensions
of the matrices to work with (for example, the even-even part of the spectrum may be obtained in the CMM using
a $900 \times 900$ matrix); second, for a general problem the matrices obtained with the CMM are not sparse,
while those obtained with the CCM are sparse. The sparsity of a matrix is a rather valuable property from
a computational point of view since it allows to save computer memory. 
If we look at the two curves in the Figure we may confirm our previous observation that the CMM provides better 
results for the lowest part of the spectrum: as a matter of fact the error for the ground state obtained with 
the CCM is of the order of $10^{-7}$. This may be understood remembering that within the collocation method
{\sl no integrals are calculated}, since all expressions are simply evaluated on the grid.

On the other hand the CCM results are obtained with a much larger matrix: for this reason we see that the error 
grows more gently for the highly excited states compared to the CMM.

\begin{figure}
\begin{center}
\bigskip\bigskip\bigskip
\includegraphics[width=7cm]{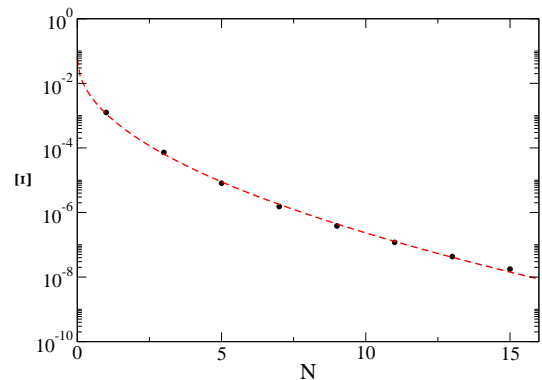}
\caption{(color online) Error over the energy of the ground state of the circular membrane 
obtained using eq.~(\ref{C8}) and minimizing the coefficients. $N$ is the cutoff in the basis
(only the wave functions $\psi_1(x)$, $\dots$, $\psi_N(x)$ are used). The line is the fit
$\Xi_{FIT} =  0.058 \times 10^{-1.7 \sqrt{N}}$.}
\label{Fig_variationalcircle}
\end{center}
\end{figure}

We now consider the application of  eq.~(\ref{C8}): we may pick the coefficients $c_n$
minimizing the expectation value in the equation, thus obtaining a precise upper bound 
to the energy of the ground state. In Fig.~\ref{Fig_variationalcircle} we plot the error over
the energy of the ground state calculated variationally. A fixed cutoff $N_{int}=36$ has been 
used for the internal sums; a varying cutoff $N$ is used to limit the number of "final" states 
$\psi_n(x)$ used in the calculation (remember that the basis in two dimensions is obtained 
with the direct product of these states and therefore grows as $N^2$).

In Table \ref{table-3} we display the energies of the twenty lowest levels of the circular membrane 
obtained using perturbation theory to different orders (up to third order). 
In the first column we report the quantum numbers corresponding to the states of a square 
box of side $2$; in the second column we report the degeneracy of the states; 
in the last column we display the exact results, obtained from  the zeroes of the Bessel functions.
The formulas obtained in this paper have been applied limiting the internal sum taking $N=20$.

There are several observations that we can make looking at this Table: first, that 
these results, although perturbative, are rather precise, expecially for the lowest 
lying states; second, that we are recovering the correct degeneracies observed in the spectrum
of the circle: in some cases, such as for the states $(1,2)$ and $(2,1)$ the 
the interaction term ($\sigma$) is diagonal and therefore the degeneracy is not lifted by
perturbation theory; in other cases, such as for the states $(1,3)$ and $(3,1)$ the interaction 
is not diagonal and therefore perturbation theory splits the levels~\footnote{As we shall see later
working with a general deformation, to leading order in PT the degeneration is lifted only for states
(we limit to twice degenerate states) $|n_x n_y\rangle$ where $n_x+n_y$ is even.}. However, the lowest of the
splitted levels approaches the level corresponding to $(2,2)$, thus approximately
reproducing the degeneracy observed in the spectrum of the circle. Finally, 
for some states we observe level crossing caused by the interaction: 
this is the case, for example, of the states $(1,5)$ and $(5,1)$,  which are splitted
by the interaction, yielding a lower state which lyies below the degenerate levels $(3,4)$ and $(4,3)$.

\begin{figure}
\begin{center}
\bigskip\bigskip\bigskip
\includegraphics[width=7cm]{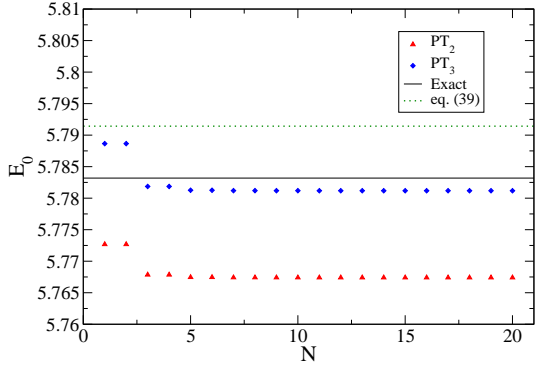}
\caption{(color online) Energy of the fundamental mode of the unit circle obtained using perturbation theory 
to order two (triangles) and three (diamonds) as a function of the number of states in the internal sums. 
The solid line is the exact result. The dotted line is the result of eq.~(\ref{C12}).}
\label{Fig_ptcircle}
\end{center}
\end{figure}

In Fig.~\ref{Fig_ptcircle} we display the energy of the fundamental mode  of the unit circle obtained 
using perturbation theory to order two (triangles) and three (diamonds) as a function of the number of
states in the internal sums. The solid line is the exact result, while the dotted line is 
the result obtained using the simple formula of eq.~(\ref{C12}).

\subsection{Deformation of the square}
\label{squaredeform}

\begin{figure}
\begin{center}
\bigskip\bigskip\bigskip
\includegraphics[width=5cm]{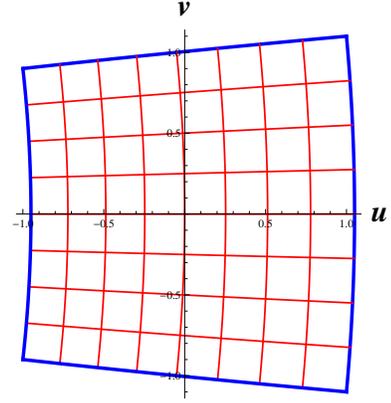}
\caption{(color online) Deformation of the square of side $2$ produced by the map $f(z) = z + z^2/20$.}
\label{Fig_2}
\end{center}
\end{figure}

We consider the conformal map $f(z) = z + \alpha z^2$, for $|\alpha| \ll 1$, which produces a deformation
of the square of side $2$. Fig.\ref{Fig_2} shows the membrane obtained for $\alpha = 1/20$.

In this case we have
\beq
\sigma(x,y) &=& 4 \alpha  x + 4 \alpha ^2 \left(x^2+y^2\right) \ .
\eeq

We may easily calculate the matrix elements using the integrals earlier defined:
\beq
\langle n | \sigma | m \rangle &=& 4 \alpha \mathcal{Q}_{n_x m_x 1} \ \delta_{n_y m_y} \nonumber \\
&+& 4 \alpha^2 \left(\mathcal{Q}_{n_x m_x 2} \ \delta_{n_y m_y} 
+\mathcal{Q}_{n_y m_y 2} \ \delta_{n_x m_x} \right) .
\eeq

In particular:
\beq
\langle n | \sigma | n \rangle &=& 4 \alpha^2 \left(\frac{2}{3}-\frac{2}{\pi ^2 n_x^2}-\frac{2}{\pi ^2 n_y^2}
\right) 
\eeq
for $n=m$ and
\beq
\langle n | \sigma | m \rangle &=& 
\alpha \frac{32 m_x n_x \left((-1)^{m_x+n_x}-1\right)}{\pi ^2 \left(m_x^2-n_x^2\right)^2}  \ \delta_{n_y m_y} 
\nonumber \\
&+& \alpha^2 \left[\frac{64 m_y n_y \left((-1)^{m_y+n_y}+1\right)}{\pi^2 \left(m_y^2-n_y^2\right)^2} \ \delta_{n_x m_x} \right. \nonumber \\
&+& \left.
\frac{64 m_x n_x \left((-1)^{m_x+n_x}+1\right)}{\pi^2 \left(m_x^2-n_x^2\right)^2} \ \delta_{n_y m_y} 
\right]
\eeq
for $n\neq m$. Let us discuss explicitly the case of states which are twice degenerate: 
$|\tilde{n}_1\rangle = |n_x,n_y\rangle$ and $|\tilde{n}_2\rangle = |n_y,n_x\rangle$ where $n_x \neq n_y$. 

For such states we have that the matrix elements read:
\beq
\langle \tilde{n}_1 | \sigma | \tilde{n}_1 \rangle &=& \langle \tilde{n}_2 | \sigma | \tilde{n}_2 \rangle = 
\frac{8}{3} \alpha^2 \left( 1- \frac{3}{\pi^2 n_x^2} - \frac{3}{\pi^2 n_y^2}\right) \nonumber \\
\langle \tilde{n}_1 | \sigma | \tilde{n}_2 \rangle &=& \langle \tilde{n}_2 | \sigma | \tilde{n}_1 \rangle = 0 \nonumber \ .
\eeq

Since the matrix is diagonal and proportional to the $2 \times 2$ identity matrix, the perturbation in this case does not
separate the degenerate states and the first order correction in PT is the same as for a the non-degenerate states:
\beq
E_n^{(1)}=- \epsilon_{n} \frac{8}{3} \alpha^2 \left( 1- \frac{3}{\pi^2 n_x^2} - \frac{3}{\pi^2 n_y^2}\right) \ .
\eeq

We notice however that $E_n^{(1)}$ goes like $\alpha^2$, so we need to calculate the $\alpha^2$ contribution which 
comes from the second order correction. As we can see from the matrix elements of $\sigma$ written above, 
to order $\alpha$ only nondiagonal terms contribute. Therefore, provided that $n_x \neq m_x$ we have:
\beq
\langle n | \sigma | m \rangle &\approx&  \alpha \frac{32 m_x n_x \left((-1)^{m_x+n_x}-1\right)}{\pi ^2 
\left(m_x^2-n_x^2\right)^2}  \ \delta_{n_y m_y} + O\left[\alpha^2\right]
\eeq

We may thus write the contribution to order $\alpha^2$ in the second order PT:
\beq
E_n^{(2)} &\approx& \alpha^2 \epsilon_n^2 n_x^2 \frac{2^{14}}{\pi^6} \ F(n_x) 
\eeq
where
\beq
F(n_x) \equiv \sum_{m_x\neq n_x} \left[ \frac{m_x^2}{2} \ 
\frac{\left((-1)^{m_x+n_x}-1\right) }{\left(n_x^2-m_x^2\right)^5} \right] \ .
\eeq

Actually these formulas hold also for states of higher degeneracy, since also in this case the 
interaction matrix is diagonal and proportional to the identity matrix.

The series appearing in this equation may be calculated analytically (we do not report 
the expression here because of its length), instead we report the first three values are:
\beq 
F(1) &=& \frac{\pi ^4}{3072}-\frac{5 \pi ^2}{1024} \approx -0.0164827 \nonumber \\
F(2) &=& \frac{\pi ^4}{49152}-\frac{5 \pi ^2}{65536} \approx 0.0012288 \nonumber \\
F(3) &=& \frac{\pi ^4}{248832}-\frac{5 \pi ^2}{746496} \approx 0.000325359 \nonumber \ .
\eeq

We may also calculate the leading behaviour of $F(n_x)$ for $n_x \rightarrow \infty$. 
In this case we find:
\beq
\lim_{n_x\rightarrow \infty} F(n_x) = \frac{\pi^4}{3072 \ n_x^4} \approx \frac{0.0317087}{n_x^4} \ .
\eeq

Notice that in the limit of highly excited levels, $n_x \rightarrow \infty$, we may use the asymptotic
behavior of $F(n_x)$ written above to predict that the energy calculated to order $\alpha^2$ 
behave as
\beq
E_n &\approx& \epsilon_n \left[1+\frac{4}{3} \alpha ^2 \left(\frac{{n_y}^2+\frac{6}{\pi ^2}}{{n_x}^2}
+\frac{6}{\pi ^2 {n_y}^2}-1\right) \right] \ .
\eeq

Provided that we consider states with $n_y \ll n_x$ we obtain the following limit
\beq
E_n &\approx& \epsilon_n \left[ 1 - \frac{4}{3} \alpha^2  \right] 
\eeq
thus being decreased {\sl by a constant factor} with respect to the energies of the box. 

In Table \ref{table-pt}  we display the quantity $(E_n - E_n^{box})/\alpha^2$,
for different values of $\alpha$ and for the first $50$ states. The energies
in the deformed box have been obtained using CCM with $N=60$. In the last
column we report the leading perturbative result, corresponding to the 
analytical formula which has been obtained in this paper. 
Notice how well the numerical results approach the theoretical 
one even for moderate values of $\alpha$. Only $15$ of these corrections are 
positive, the remaining corresponding to a lowering of the frequencies.

\begin{figure}
\begin{center}
\bigskip\bigskip\bigskip
\includegraphics[width=8cm]{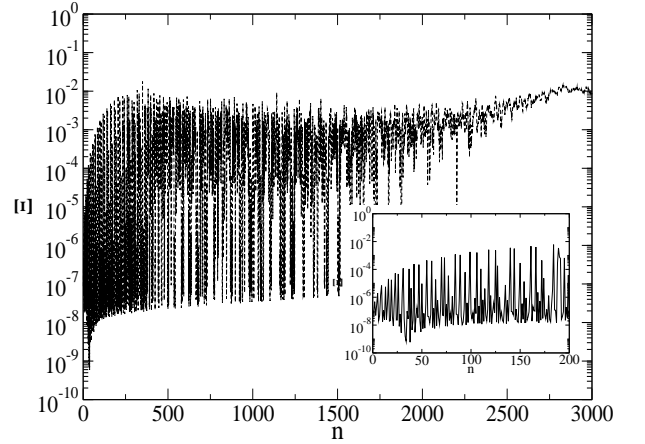}
\caption{$\Xi \equiv \left| 1 - \frac{E_n^{(CCM)}}{E_n^{(PT)}} \right|$ for $\alpha = 1/100$
for the first 3000 levels. The numerical CCM results have been obtained with a grid corresponding to
$N=100$. The inset plot is the blow-up of the first 200 levels.}
\label{Fig_spectrum}
\end{center}
\end{figure}

Fig.~\ref{Fig_spectrum} displays the quantity $\Xi \equiv \left| 1 - \frac{E_n^{(CCM)}}{E_n^{(PT)}} \right|$ 
for $\alpha = 1/100$ for the first 3000 levels.  The numerical CCM results have been obtained with a grid 
corresponding to $N=100$. The inset plot is the blow-up of the first 200 levels.

\begin{figure}
\begin{center}
\bigskip\bigskip\bigskip
\includegraphics[width=8cm]{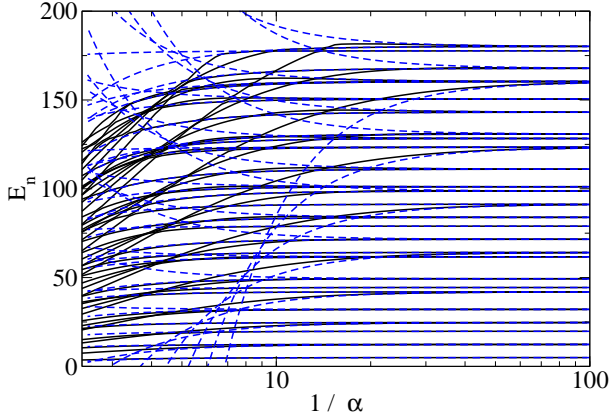}
\caption{(color online) First 50 levels of the deformed square as a function of $1/\alpha$. The solid lines 
are the numerical results obtained with CCM with $N=60$; the dashed lines are the analytical perturbative 
results obtained in this paper.}
\label{Fig_spectrum2}
\end{center}
\end{figure}

In Fig.~\ref{Fig_spectrum2} we display the first $50$ levels of the deformed square as a function
of $1/\alpha$. The solid line are the precise numerical results obtained using CCM with $N=60$;
the dashed lines are obtained using the analytical formulas that we have obtained.

\subsection{A general formula}
\label{hear}

Let us therefore consider a general conformal map
\beq
f(z) = z + \eta \sum_{n=1}^\infty \frac{\varrho_n}{n} z^n \ ,
\eeq
where $\varrho_n$ are real coefficients of similar strength and $\eta$ is a small parameter.
Notice that $\varrho_1$ is related to a dilatation.

In this case we have
\beq
\sigma &=& \eta \sum_{n=2}^{\infty} \varrho_n \left( (x+i y)^{n-1}+ (x-i y)^{n-1} \right) \nonumber \\
&+& \eta^2 \sum_{n,m=2}^{\infty} \varrho_n \varrho_m (x+i y)^{n-1} (x+i y)^{m-1} \nonumber \\
&=& \eta \sum_{n=2}^{\infty} \varrho_n \sum_{k=0}^{n-1} \ \left(\begin{array}{c}
n-1\\
k\\
\end{array}
\right) \ x^k y^{n-1-k} i^{n-1-k}\nonumber \\
&\times& \left(1+ (-1)^{n-1-k}\right) \nonumber \\
&+& \eta^2 \sum_{n,m=2}^{\infty} \varrho_n \varrho_m (x+i y)^{n-1} (x+i y)^{m-1} \ .
\eeq

We focus on the term of order $\eta$ and write the corresponding matrix elements
\beq
\langle n | \sigma | m \rangle &\approx& \eta \sum_{n=2}^{\infty} \varrho_n C_{n ; n_x m_x n_y  m_y} + 
O\left[\eta^2\right] \ ,
\eeq
where 
\beq
 C_{n ; n_x m_x n_y  m_y} &\equiv&
\sum_{k=0}^{n-1} \ \left(\begin{array}{c}
n-1\\
k\\
\end{array}
\right) \ \mathcal{Q}_{n_x m_x k} \mathcal{Q}_{n_y m_y n-1-k} \nonumber\\
&& i^{n-1-k} \left(1+ (-1)^{n-1-k}\right) \ .
\eeq
 
Notice that an analytical expression for the $C$ is obtained quite easily using the results in the appendix \ref{app_a}. In particular
the diagonal matrix elements read
\begin{widetext}
\beq
\langle n | \sigma | n \rangle &\approx& \eta \left\{
\varrho_3 \ \frac{4 \left(\frac{1}{n_y^2}-\frac{1}{n_x^2}\right)}{\pi ^2} + \varrho_5 \left(
\frac{48 \left(n_x^4-n_x^2 n_y^2+n_y^4\right)}{\pi ^4 n_x^4 n_y^4}-\frac{8}{15}\right) \right. \nonumber \\
&-& \left.\varrho_7 \left(\frac{16 (n_x-n_y) (n_x+n_y) \left(\pi ^4 n_x^4 n_y^4-90 \left(n_x^4+
n_y^4\right)\right)}{\pi ^6 n_x^6 n_y^6}\right) \right. \nonumber \\
&+& \left. \varrho_9
\left(\frac{896 \left(90 \left(n_x^8-n_x^6 n_y^2+n_x^4 n_y^4-n_x^2 n_y^6+n_y^8\right)-\pi^4 n_x^4
   n_y^4 \left(n_x^4-n_x^2 n_y^2+n_y^4\right)\right)}{\pi^8 n_x^8 n_y^8}+\frac{32}{45}\right)
+ \dots\right\} \ .
\eeq
\end{widetext}

We may now consider the off-diagonal matrix elements between two degenerate states. For the twice degenerate states
$| n \rangle = |1,2\rangle$ and $| m \rangle = |2,1\rangle$ we have
\beq
\langle 12 | \sigma | 21 \rangle &=& \langle 21 | \sigma | 12 \rangle = 0 \ .
\eeq
In this case the degeneration of the levels is not lifted by the perturbation as already discussed earlier. The same
occurrs for all the states where $n_x$ and $n_y$ add to an odd integer.

On the other hand, states where  $n_x$ and $n_y$ add to an even integer do have non vanishing matrix elements. For example,
for the states $| n \rangle = |1,3\rangle$ and $| m \rangle = |3,1\rangle$  we have:
\begin{widetext}
\beq
\langle 13 | \sigma | 31 \rangle &=& \langle 31 | \sigma | 13 \rangle = -\frac{27 \varrho_5}{\pi^4}
+\frac{63 \left(8 \pi^4-765\right) \varrho_9}{\pi^8}-\frac{297 \left(3869775-42840 \pi^4+32 \pi^8\right) \varrho_{13}}{2 \pi^{12}} \nonumber \\
&+&\frac{135 \left(-185977516725+2065943880 \pi^4-1633632 \pi^8+256 \pi^{12}\right) \varrho_{17}}{\pi^{16}} + \dots
\eeq
\end{widetext}

For these states the degeneracy is lifted by the perturbation.

In Table \ref{table-4} we have written the analytical expressions for the first order perturbative corrections (divided by
the energy of the square of corresponding quantum numbers and changed of sign) generated by the conformal map  written above,
for the first $20$ energy levels. The reader may notice that only odd powers in the map contribute to this order.
Moreover, the degeneration is lifted only for the states $\left\{ |1,3\rangle , |3,1\rangle\right\}$, $\left\{ |2,4\rangle , |4,2\rangle\right\}$ and
$\left\{ |1,5\rangle , |5,1\rangle\right\}$ (for which as mentioned before the sum of the two quantum numbers is even), 
although the splittings depend only on some of the coefficients ($\varrho_5$,$\varrho_9$, $\dots$).

\subsection{Robnik's billiards}     

Out next application is to Robnik's billiard, which are  obtained using  the conformal map of the circle~\cite{Robnik84}
\beq
f(z) &=& A z + B z^2 \ ,
\eeq
where $A=\cos p$ and $B=\frac{1}{\sqrt{2}} \sin p$. The parametrization of $A$ and $B$ in terms of $p$ describes a family
of billiards of constant area $A=\pi$. We write the map as
\beq
f(z) &=& \cos p \ \left[ z + \lambda z^2\right]  \equiv \cos p \ \bar{f}(z) \ .
\eeq 
where we have defined $\lambda \equiv \frac{\tan p}{\sqrt{2}}$ as in Ref.~\cite{Robnik84}.
Since the maps $f(z)$ and $\bar{f}(z)$ are related by a dilatation, one can calculate the eigenvalues 
corresponding to the second map and easily obtain the eigenvalues corresponding to the first map as:
\beq
E_n = \frac{\bar{E}_n}{\cos^2 p} \ .
\eeq

\begin{figure}
\begin{center}
\bigskip\bigskip\bigskip
\includegraphics[width=5cm]{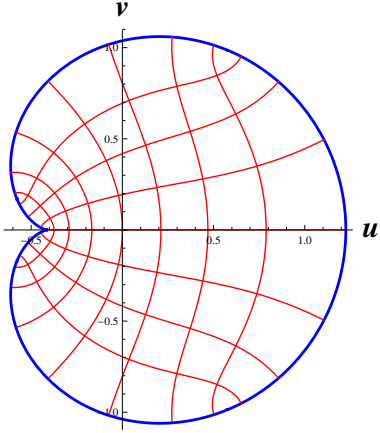}
\caption{(color online) Robnik's billiard for $\lambda=1/2$.}
\label{Fig_robnik}
\end{center}
\end{figure}

We will therefore use the second map and then use the relation above to obtain the eigenvalues corresponding
to the first map. We thus have:
\beq
\Sigma = 1 + 4 \lambda r \cos\theta + 4 \lambda^2 r^2 \ ,  
\eeq
and
\beq
\sigma =  4 \lambda r \cos\theta + 4 \lambda^2 r^2 \ .
\eeq

Notice that having chosen to work with the second map rather than with the first one, eliminates a constant
term $\sigma$ and therefore simplifies the calculation.

In this case we are working with the orthonormal basis of the circle:
\beq
\phi_{kns}(r,\theta) \equiv R_{kn} J_k(\gamma_{kn} r) \times 
\left\{ \begin{array}{cc} \cos k \theta  & , \ s=1 \\ \sin k\theta & , \ s=2\\
\end{array}\right.
\eeq

The normalization constants are:
\beq
R_{0n} &=& \frac{1}{\sqrt{\pi}\ J_0'(\gamma_{0n} )} \ \ , \ k = 0 \\
R_{kn} &=& \frac{\sqrt{2}}{\sqrt{\pi}\ J_k'(\gamma_{kn} )} \ \ , \ k > 0 ,
\eeq
where $J_k$ is the Bessel function of order $k$ and $\gamma_{kn}$ its zeroes. 
We have followed almost entirely the notation of Ref.~\cite{Robnik84}, apart of introducing an index $s$ for 
the degeneracy: all states with $k>0$ are twice degenerate, states with $k=0$ are simple.

The application of our perturbative formula requires the calculation of the matrix elements of $\sigma$. 
Thus we need to calculate:
\begin{widetext}
\beq
\langle kns | r^2 | k'n's'\rangle &=& \int_0^1 dr \int_0^{2\pi} d\phi \ r^3 \phi_{kns}(r,\theta)\phi_{k'n's'}(r,\theta) 
\nonumber \\
&=& \pi \left[ \delta_{k,k'} + (-1)^{s+1} \delta_{k,-k'} \right] \delta_{ss'} R_{kns} R_{kn's}  \int_0^1 dr r^3 J_k(\gamma_{kn} r) 
J_k(\gamma_{kn'} r) \\
\langle kns | r \cos\theta | k'n's'\rangle &=& \int_0^1 dr \int_0^{2\pi} d\phi \ r^2 \cos\theta \phi_{kns}(r,\theta)
\phi_{k'n's'}(r,\theta) \nonumber \\
&=& g_{kk's} \delta_{ss'} R_{kns} R_{k'n's}  \int_0^1 dr r^2 J_k(\gamma_{kn} r) J_k(\gamma_{kn'} r)
\eeq
\end{widetext}
where 
\beq
g_{kk's} &=& \frac{\pi}{2} \ \left[ \left(\delta_{k-k',1} + \delta_{k-k',-1} \right) \right. \nonumber \\
&+& \left. (-1)^{s+1}  \left(\delta_{k+k',1} + \delta_{k+k',-1} \right) \right]   \ .
\eeq

Using the integrals above we may obtain the matrix elements of $\sigma$:
\beq
\langle kns | \sigma | k' n's\rangle &\equiv& \lambda \langle kns | \sigma^{(1)} | k' n's\rangle +
\lambda^2 \langle kns | \sigma^{(2)} | k' n's\rangle
\nonumber \\
&=&  4 \lambda \langle kns | r \cos\theta | k' n's\rangle  \nonumber \\
&+& 4 \lambda^2 \langle kns | r^2 | kn's\rangle \ .
\eeq

Notice that the diagonal matrix elements are of order $\lambda^2$, whereas the off-diagonal matrix elements start at
order $\lambda$. This means that our perturbative (in the power counting parameter $\eta$) formulas contain 
contributions with different orders in $\lambda$. For example the first order formula is of order $\lambda^2$,
just as the second order perturbative formula, which contains squares of off-diagonal matrix elements.
On the other hand, the third order perturbative formula is of order $\lambda^4$, since the last sum, which involves
the product of three non-diagonal matrix elements vanishes when the external states are taken to be equal.
For this reason, the general results that we have obtained in this paper allow us to make a rigorous 
perturbative analysis of this problem only to order $\lambda^2$.

We may therefore write
\beq
\bar{E}_{kns} = \bar{E}_{kns}^{(0)} + \lambda^2 \ \bar{E}_{kns}^{(2)} + O\left[\lambda^4\right]
\label{ptrob}
\eeq
where
\beq
\bar{E}_{kns}^{(0)} &=& \gamma_{kn}^2 \nonumber \\
\bar{E}_{kns}^{(2)} &=& -\gamma_{kn}^2 \langle kns | \sigma^{(2)} | k n s \rangle + \gamma_{kn}^4
\sum'_{k',n'} \frac{\langle kns | \sigma^{(1)} | k' n' s \rangle^2}{\gamma_{kn}^2 -\gamma_{k'n'}^2 }  \nonumber
\eeq
and $\sum'_{k',n'}$ means the sum obtained excluding the values $(k',n')=(k,n)$.

Even before doing any numerical calculation, we may look at the form of the matrix elements of $\sigma$ to
understand how the degeneracy of the level of the circle is lifted. We observe that the $s$ dependent
component of $g_{k,k',s}$ is nonzero only when $(k,k')$ is $(0,1)$ or $(1,0)$. This means that
only the levels with principal quantum number $k=1$ have the degeneracy lifted. A splitting
of the levels corresponding to $k>1$ may only occurr through higher order perturbative contributions (to 
order $\lambda^4$ and higher, and it is therefore highly suppressed.

In Table \ref{table-5} we display the first $40$ eigenvalues of a Robnik billiard with $\lambda = 1/100$ and
$\lambda=1/20$, calculated with the perturbative formula (\ref{ptrob}) to order $\lambda^0$ and 
$\lambda^2$ and with the CCM with $N=100$. We may observe that the splitting of the levels corresponding
to $k=1$ is reproduced quite well by our formula; in particular the results obtained for $\lambda=1/100$ 
with second order perturbation theory reproduce amazingly well the numerical CCM results. For $\lambda=1/20$,
we observe a small splitting of some of the levels corresponding to $k>1$, such as for example for the
states with quantum numbers $(2,1)$. As mentioned before, these splittings come from higher order
perturbative corrections, which we are not considering here.

\begin{figure}
\begin{center}
\bigskip\bigskip\bigskip
\includegraphics[width=8cm]{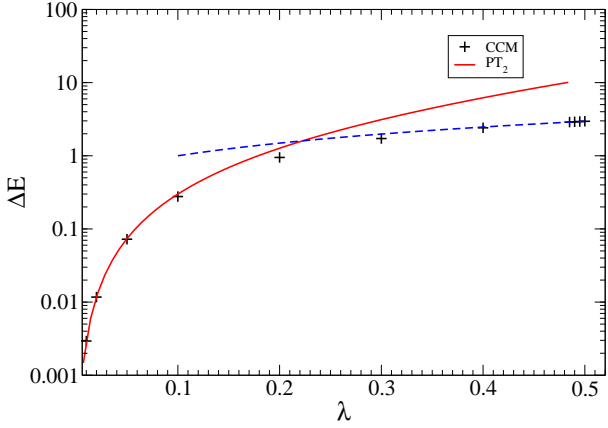}
\caption{(color online) Splitting of the levels corresponding to $(k,n) = (1,1)$. The solid line is 
the perturbative result to order $\lambda^2$; the pluses are the numerical results obtained with 
CCM with $N=100$. The dashed line is the linear fit of the four last numerical values 
$\Delta E^{(fit)} = 0.5095 + 4.90231 \lambda$.}
\label{Fig_DeltaE}
\end{center}
\end{figure}

In Fig.~\ref{Fig_DeltaE} we display the splitting of the first two excited levels of the Robnik's billiard,
corresponding to the quantum numbers $(k,n)=(1,1)$, as a function of the parameter $\lambda$. The solid
line is the perturbative result to order $\lambda^2$, calculated in this paper, which corresponds to
$\Delta E^{(PT)} = 29.3639418 \lambda^2 + 58.7278836 \lambda^4$ (observe that this expression contains 
a power $\lambda^4$ because of extra power of $\lambda$ coming from the rescaling factor $1/\cos^2 p$ 
in the energies); the pluses are the 
precise numerical results obtained using CCM with $N=100$. Finally, for $\lambda \approx 0.5$, the
splitting is seen to behave linearly in $\lambda$: the dashed line represent the linear fit of the
last four points, $\Delta E^{(fit)} = 0.5095 + 4.90231 \lambda$. A resummation of the 
perturbative series (assuming that it is possible to calculate it to arbitrary orders), 
should therefore be able to reproduce this leading asymptotic behaviour. It would also be interesting to 
study the radius of convergence of the perturbative series and see if it has any relation with the
insurgence of quantum chaos.

\subsection{Regular polygonal billiard}

Another interesting application of the techniques discussed in this paper is to the calculation of the frequencies of 
a regular polygonal billiard. This problem has been attacked in a series of papers, using both numerical and analytical
approaches. 

In ref.~\cite{Liboff94} Liboff has studied the polygonal quantum billiard problem, proving that
the first excited state of a regular polygon is doubly degenerate. In ref.~\cite{Liboff01a} Liboff and Greenberg
have studied the hexagon quantum billiard, constructing a subset of eigenfunctions and eigenvalues, 
while in ref.~\cite{Liboff01b} Liboff has considered classical and quantum wedge billiards.

Molinari has obtained analytical expressions for the frequency of the ground and first 
excited states using a perturbative approach, which he calls "lambda expansion"~\cite{Molinari97}. 
Cureton and Kuttler~\cite{Kuttler99} have used a numerical approach (which is essentially the Conformal 
Mapping Method of \cite{Robnik84}), to obtain quite precise values for selected eigenvalues of regular polygons 
and of the figures obtained by the dissection.
More recently, Grinfield and Strang~\cite{Grinfield04} have obtained an analytical formula for the simple 
states of regular polygonal with ${\cal N}$ sides as a function of $1/{\cal N}$. 

Other works have considered specific polygons, and applied different methods to solve the Helmholtz equation on these 
domains:  for example, Betcke and  Trefethen have applied the method of particular solutions in ref.~\cite{Betcke05} 
to obtain very precise values for the first three eigenvalues of a regular decagon; Lijnen, Chibotaru and
Ceulemans, used a radial rescaling approach in ref.~\cite{Lijnen08} to obtain the ten lowest eigenvalues
of regular polygons with ${\cal N}=3, \dots, 6$ sides; finally, Guidotti and Lambers~\cite{Guidotti08} have 
considered a variant of the method of particular solution, and applied to a regular polygon with $128$ sides 
(comparing their findings with those predicted by the formula of Grinfield and Strang).

We will now treat this problem with the systematic perturbative approach developed in this paper and then 
perform a comparison with the works mentioned above. 

Our starting point is the conformal map
\beq
f(z) = C_{\cal N} \int_0^z \frac{ds}{(1-s^{\cal N})^{2/{\cal N}}} \  ,
\eeq
which maps the unit circle into a regular polygon of $N$ sides inscribed in the circle~\cite{Molinari97,Kuttler99}. 
Following Molinari we have defined 
$C_{\cal N} \equiv  \frac{\Gamma(1-1/{\cal N})}{\Gamma(1+1/{\cal N}) \Gamma(1-2/{\cal N})}$ and
we have expanded the map as
\beq
f(z) =  C_{\cal N} \sum_{k=0}^\infty f_k z^{{\cal N} k+1} 
\eeq
where $f_0=1$ and
\beq
f_k = \frac{1}{k! ({\cal N} k+1)} \ \prod_{j=0}^{k-1} \left(\frac{2}{\cal N}+j\right) 
\eeq
for $k\geq 1$.

As explained in the previous section, it is convenient to work with a rescaled map defined 
through the relation $f(z) \equiv C_{\cal N} \bar{f}(z)$ and then easily relate the eigenvalues
corresponding to the map $\bar{f}(z)$ to those corresponding to the map $f(z)$ by means of the
equation
\beq
E_n = \frac{\bar{E}_n}{C_{\cal N}^2} \ .
\label{eq:polygon}
\eeq

We thus obtain
\beq
\Sigma &=& \sum_{k=0}^\infty \sum_{j=0}^\infty  ({\cal N} k+1) ({\cal N} j+1) f_k f_j r^{{\cal N} (k+j)} 
e^{i {\cal N} (k-j) \theta} \nonumber 
\eeq
and
\beq
\sigma &=& -1 +\sum_{k=0}^\infty \sum_{j=0}^\infty  ({\cal N} k+1) ({\cal N} j+1) f_k f_j r^{{\cal N} (k+j)} 
e^{i {\cal N} (k-j) \theta} \nonumber \ .
\eeq

Let us calculate explicitly the matrix elements $\langle kns | \sigma | k'n's\rangle$.
We define
\beq
{\cal I}^{(1)}_{k,k',l,m} &\equiv& \int_0^{2\pi} \cos k\theta \ \cos k'\theta \ \cos {\cal N}(l-m)\theta \ 
d\theta \nonumber \\
&=& \frac{\pi}{2} \left[ \delta_{k-k'+{\cal N}(l-m)} + \delta_{k+k'+{\cal N}(l-m)} \right.\nonumber \\
&+& \left. \delta_{k-k'-{\cal N}(l-m)} + \delta_{k+k'-{\cal N}(l-m)} \right] \\
{\cal I}^{(2)}_{k,k',l,m} &\equiv& \int_0^{2\pi} \sin k\theta \ \sin k'\theta \ \cos {\cal N}(l-m)\theta \ d\theta \nonumber \\
&=& \frac{\pi}{2} \left[ \delta_{k-k'+{\cal N}(l-m)} - \delta_{k+k'+{\cal N}(l-m)} \right.\nonumber \\
&+& \left. \delta_{k-k'-{\cal N}(l-m)} - \delta_{k+k'-{\cal N}(l-m)} \right] 
\eeq
which can be written as
\beq
{\cal I}^{(s)}_{k,k',l,m} &\equiv&  {\cal I}^{(a)}_{k,k',l,m} + (-1)^{s+1} {\cal I}^{(b)}_{k,k',l,m}\nonumber \\
&=&\frac{\pi}{2} \left[ \delta_{k-k'+{\cal N}(l-m)} + \delta_{k-k'-{\cal N}(l-m)} \right] \nonumber \\
&+& (-1)^{s+1} \frac{\pi}{2} \left[ \delta_{k+k'+{\cal N}(l-m)} + \delta_{k+k'-{\cal N}(l-m)} \right]  \nonumber \ .
\eeq
We also define
\beq
{\cal J}_{kn,k'n',{\cal N},l,m} \equiv \int_0^1  r^{{\cal N} (l+m) +1} J_{k}(\gamma_{kn}r) J_{k'}(\gamma_{k'n'}r) dr \nonumber
\eeq
and thus we may write the matrix element of $\sigma$ as:
\begin{widetext}
\beq
\langle kns | \sigma | k'n's'\rangle &=& - \delta_{kk'} \delta_{nn'} \delta_{ss'} +
\sum_{l=0}^\infty \sum_{m=0}^\infty  ({\cal N} l+1) ({\cal N} m+1) f_l f_m R_{kn} R_{k'n'} \delta_{ss'} 
{\cal I}^{(s)}_{k,k',l,m} {\cal J}_{kn,k'n',{\cal N},l,m} \ .
\eeq
\end{widetext}

We focus here on the calculation of the first order correction to the eigenvalues:
\beq
\bar{E}_{kns}^{(1)} &=& - \gamma_{kn}^2 \langle kns | \sigma | kns\rangle \ .
\eeq

We may now explicitly write the diagonal matrix element of $\sigma$:
\begin{widetext}
\beq
\langle kns | \sigma | kns\rangle &=& \sum_{l=1}^\infty   ({\cal N} l+1)^2 f_l^2 R_{kn}^2 {\cal I}^{(a)}_{k,k,l,l} 
{\cal J}_{kn,kn,{\cal N},l,l} \nonumber \\
&+& (-1)^{s+1} \sum_{l=0}^\infty \sum_{m=0}^\infty  ({\cal N} l+1) ({\cal N} m+1) f_l f_m R_{kn}^2   {\cal I}^{(b)}_{k,k,l,m} 
{\cal J}_{kn,kn,{\cal N},l,m} \ ,
\eeq
\end{widetext}
where the last term, when non-zero, breaks the degeneracy of the states with $s=1$ and $s=2$. If we look at the 
form of ${\cal I}^{(b)}_{k,k,l,m}$ we see that it does not vanish only when the condition $2 k = \pm {\cal N} (l-m)$ is met.
We need to distinguish between regular polygons with even or odd numbers of sides:
for ${\cal N}$ even, we may fulfill this condition for $k={\cal N}/2$ and $l=m \pm 1$, for $k={\cal N}$ and $l=m \pm 2$, and so on;
for ${\cal N}$ odd, this condition is met for $k={\cal N}$ and $l=m \pm 2$, for $k= 2{\cal N}$ and $l=m \pm 4$, and so on. 
Thus, perturbation theory to first order predicts a peculiar pattern of lifting of the degeneracy of levels.

Notice that the radial integrals can also be calculated explicitly:
\begin{widetext}
\beq
{\cal J}_{kn,kn,{\cal N},l,m} &=& \frac{4^{-k} \gamma_{k,n}^{2 k}}{\Gamma (k+1)^2 (2 k+{\cal N}
   (l+m)+2)} \nonumber \\
&\times&  _2F_3\left(k+\frac{1}{2},k+\frac{l {\cal N}}{2}+\frac{m {\cal N}}{2}+1;k+1,2
   k+1,k+\frac{l {\cal N}}{2}+\frac{m {\cal N}}{2}+2;-\gamma_{k,n}^2\right) \nonumber \ .
\eeq 
\end{widetext}

We can therefore write the energy calculated up to first order in perturbation theory as
\beq
E_{kns} =   \frac{\gamma_{kn}^2}{C_{\cal N}^2} \left[ 1 -  \langle kns | \sigma | kns\rangle \right] \ .
\eeq

Let us focus for a moment on the perturbative result obtained at order zero, $E_{kns} \approx   \frac{\gamma_{kn}^2}{C_{\cal N}^2}$.
We may expand the factor $1/C_{\cal N}^2$ around ${\cal N}=\infty$, obtaining the energies
\beq
\frac{\gamma_{kn}^2}{C_{\cal N}^2} &\approx& \gamma_{kn}^2 \left[1 + \frac{2\pi^2}{3 {\cal N}^2} +   \frac{4 \zeta(3)}{{\cal N}^3} + \frac{14 \pi ^4}{45 {\cal N}^4} +\dots \right] \nonumber \ ,
\eeq
which reproduce the formula obtained by Grinfield and Strang in \cite{Grinfield04} up to order $1/{\cal N}^2$ (notice 
however that their formula is limited to the simple states). The coefficient of order $1/{\cal N}^3$ is 
different, although very similar numerically: we obtain $4\zeta(3) \approx 4.81$, while 
ref.\cite{Grinfield04} quotes $\pi^3/6 \approx 5.2$~\footnote{The value found by Grinfield and Strang is 
based on a conjecture based on numerical evidence.}.
On the other hand, if we look now at the first order correction, we clearly see that it is not analytical at
${\cal N}=\infty$ and therefore that it cannot be expanded in powers of $1/{\cal N}$. 

\begin{figure}
\begin{center}
\bigskip\bigskip\bigskip
\includegraphics[width=8cm]{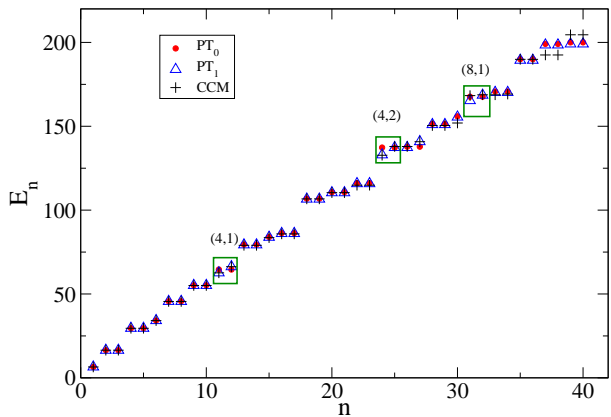}
\caption{(color online) First 40 levels of a octagon billiard. The circles (triangles) are the results obtained with
zero (first) order perturbation theory; the pluses are the exact (numerical) results obtained using CCM with $N=100$. 
The boxes enclose the levels where first order perturbation theory predicts that the degeneracy of levels is lifted.
The corresponding quantum numbers are displayed.}
\label{Fig_octagon}
\end{center}
\end{figure}

\begin{figure}
\begin{center}
\bigskip\bigskip\bigskip
\includegraphics[width=8cm]{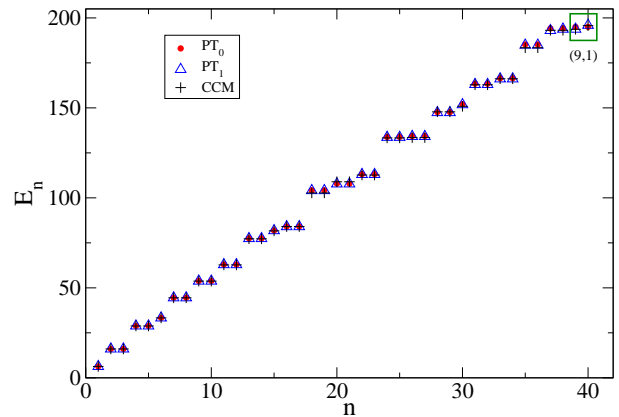}
\caption{(color online) Same as for Fig.~\ref{Fig_octagon} for the nonagon billiard.}
\label{Fig_nonagon}
\end{center}
\end{figure}

\begin{figure}
\begin{center}
\bigskip\bigskip\bigskip
\includegraphics[width=8cm]{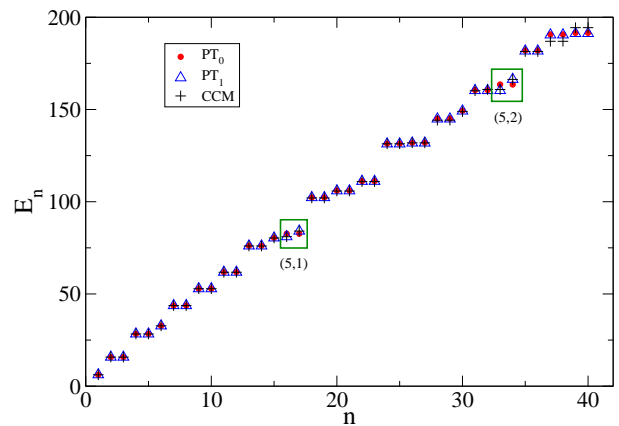}
\caption{(color online) Same as for Fig.~\ref{Fig_octagon} for the decagon billiard.}
\label{Fig_decagon}
\end{center}
\end{figure}

In Fig.~\ref{Fig_octagon}, \ref{Fig_nonagon} and \ref{Fig_decagon} we have plotted the first $40$ levels of 
the octagon, nonagon and decagon billiards, obtained using zero and first order perturbation theory 
(circles and triangle respectively) and using CCM with a grid corresponding to $N=100$ (pluses). 
The boxes in the plots enclose the levels where first order PT predicts that degeneracy is lifted: as
we have already discussed before, this occurrs at $k={\cal N}/2$ (and multiple integers) for regular polygons 
with even number of sides ${\cal N}$ and at $k = {\cal N}$ (and multiple integers) for regular polygons with odd number of sides.
The numerical values corresponding to these plots are reported in Table \ref{table-6}: we like to underline that
in some cases the CCM may provide tiny splittings for states which are degenerate. This occurrs because
the discretization may explicitly break the symmetry of the billiard, although this splitting is bound to 
decrease as the grid gets finer and finer. Taking into account this fact, and looking at the Table 
we may speculate that the states lying below the $({\cal N}/2,1)$ states for even billiards and $({\cal N},1)$ for odd billiards
are degenerate (remember that Liboff's theorem of ref.~\cite{Liboff94} tells us that the first excited
state of the regular polygonal billiard is degenerate). A rigorous proof of this conjecture would require the
calculation of higher order perturbative contributions, which we plan to consider in the future.

\newpage

\subsection{Hamiltonians with position dependent effective mass}
\label{pdem}

As we have mentioned in section \ref{subsec1}, the symmetrized form of the density dependent negative
Laplacian operator that we have proposed in this paper had already been introduced in a paper by 
Zhu and Kroemer, ref.~\cite{Zhu83}, discussing a different problem. 

In recent years there has been lot of interest in discussing the solutions of quantum mechanical models 
containing density dependent  effective masses, which are common in the description of condensed matter systems. 
However most of the efforts have been put into constructing exactly and quasi-exactly solvable potentials 
for the Schr\"odinger equation with position dependent mass terms~\cite{PDEM}. To the best of our knowledge
perturbation theory has not been used as a tool to construct the solutions to this problem; since the results 
obtained earlier  in our paper apply straightforwardly to the present case we now extend the approach of
section \ref{subsec2} to Hamiltonians of the form:
\beq
\hat{H} = \frac{1}{\sqrt{\Sigma}} \left(-\frac{\hbar^2}{2m} \hat{\Delta}\right) \frac{1}{\sqrt{\Sigma}}  + V \ .
\eeq

Notice that the stationary Schr\"odinger equation associated to this operator may be solved numerical using
the CCM of section \ref{sec2}, with the difference that now $\Sigma$ is a physical density, not related 
to a conformal mapping. Moreover, the matrix representing the operator $\hat{H}$ on the uniform grid, now contains
also a diagonal term corresponding to the discretization of the potential $V$. For problems defined on whole space,
a suitable strategy corresponds to solve the SSE in a $d$-dimensional cube ($d$ being the dimension of the problem), 
whose dimensions have been determined variationally, as done in refs.~\cite{Amore07,Amore09b}. The only difference
in this case is the presence of the spatially dependent density $\Sigma$.

An alternative way of solving the problem consists of resorting to perturbation theory.
We may cast this hamiltonian into the form
\beq
\hat{H} = \frac{1}{\sqrt{\Sigma}} \left(-\frac{\hbar^2}{2m} \hat{\Delta} + V_0 \right) \frac{1}{\sqrt{\Sigma}}  + 
\left(V-\frac{V_0}{\Sigma}\right) \ ,
\eeq
where $V_0$ is a potential for which the "unperturbed" Hamiltonian $\hat{H}_0 = 
\left(-\frac{\hbar^2}{2m} \hat{\Delta} + V_0 \right)$ is exactly solvable. The analysis carried out 
in section \ref{subsec2} may now be easily generalized, by expressing $\Sigma = 1+\eta \sigma$ as before
and letting $\left(V-\frac{V_0}{\Sigma}\right) \rightarrow \eta \left(V-\frac{V_0}{1+\eta \sigma}\right)$.
Thus we have
\beq
\hat{H} = \hat{H}_0 + \eta \hat{H}_1 + \eta^2 \hat{H}_2 + \eta^3 \hat{H}_3 + \dots
\eeq
where 
\beq
\hat{H}_1 &\equiv& \hat{H}_1^{(a)} +\hat{H}_1^{(b)} \nonumber \\
&=& - \frac{1}{2} \left[ \sigma \hat{H}_0 + \hat{H}_0 \sigma \right] + V-V_0 \\
\hat{H}_2 &\equiv& \hat{H}_2^{(a)} +\hat{H}_2^{(b)} \nonumber \\
&=&  \frac{1}{8} \left[ 2 \sigma \hat{H}_0 \sigma + 3 \sigma^2 \hat{H}_0 + 3 \hat{H}_0 \sigma^2 \right] +
V_0 \sigma \\
\hat{H}_3 &\equiv& \hat{H}_3^{(a)} +\hat{H}_3^{(b)} \nonumber \\
&=& - \frac{3}{16} \left[ \sigma^2 \hat{H}_0 \sigma + \sigma \hat{H}_0 \sigma^2 \right]
- \frac{5}{16} \left[ \sigma^3 \hat{H}_0  +  \hat{H}_0 \sigma^3 \right] \nonumber \\
&-& V_0 \sigma^2 \ ,
\eeq
where the operators $\hat{H}_i^{(a)}$ are those which contain $\hat{H}_0$ and are equivalent to the
one obtained before. Working up to second order and calling
$\epsilon_n$ and $|n\rangle$ the eigenenergies and eigenstates of $\hat{H}_0$ we find
\beq
E_n^{(1)} &=& - \epsilon_n \langle n | \sigma | n \rangle + \langle n | V-V_0 | n\rangle \\ 
E_n^{(2)} &=& \epsilon_n \langle n | \sigma|n\rangle^2 + \epsilon_n^2 \sum_{k\neq n} 
\frac{ \langle n | \sigma| k\rangle^2}{\epsilon_n-\epsilon_k} \nonumber \\
&+& 2 \sum_{k \neq n} \frac{ \langle n | \hat{H}_1^{(a)} | k \rangle \langle k | \hat{H}_1^{(b)} | n \rangle
}{\epsilon_n-\epsilon_k} \nonumber \\
&+& \sum_{k \neq n} \frac{ \langle n | \hat{H}_1^{(b)} | k \rangle^2}{\epsilon_n-\epsilon_k}  \ .
\eeq

Let us consider the term:
\beq
E_n^{(2b)} &\equiv& 2 \sum_{k \neq n} \frac{ \langle n | \hat{H}_1^{(a)} | k \rangle \langle k | \hat{H}_1^{(b)} | n \rangle}{\epsilon_n -\epsilon_k}  \nonumber \\
&=& - \sum_{k\neq n} \frac{\epsilon_n + \epsilon_k}{\epsilon_n - \epsilon_k} \langle n | \sigma | k \rangle 
\langle k | V-V_0 | n \rangle \nonumber \\
&=& - \left( \langle n | \sigma (V-V_0)| n \rangle -\langle n | \sigma |n\rangle \langle n | V-V_0 | n \rangle 
\right) \nonumber \\
&-& 2  \sum_{k\neq n} \frac{\epsilon_k}{\epsilon_n - \epsilon_k} \langle n | \sigma | k \rangle 
\langle k | V-V_0 | n \rangle \ .
\eeq

Therefore we have
\beq
E_n^{(2)} &=& \epsilon_n \langle n | \sigma|n\rangle^2 + \epsilon_n^2 \sum_{k\neq n} 
\frac{ \langle n | \sigma| k\rangle^2}{\epsilon_n-\epsilon_k} \nonumber \\
&-& \left( \langle n | \sigma (V-V_0)| n \rangle -\langle n | \sigma |n\rangle \langle n | V-V_0 | n \rangle 
\right) \nonumber \\
&-& 2  \sum_{k\neq n} \frac{\epsilon_k}{\epsilon_n - \epsilon_k} \langle n | \sigma | k \rangle 
\langle k | V-V_0 | n \rangle \nonumber \\
&+& \sum_{k \neq n} \frac{ \langle n | V-V_0 | k \rangle^2}{\epsilon_n -\epsilon_k}  \ .
\eeq

Likewise we may find the perturbative corrections to the eigenstates of the Hamiltonian; for example, we
may consider the special case where $V = V_0/\sqrt{\Sigma}$, and therefore obtain to first order
in perturbation theory:
\beq
| \Psi_n \rangle = |n \rangle -\frac{\eta}{2} \sum_{k\neq n}  \frac{\epsilon_n + \epsilon_k}{\epsilon_n - \epsilon_k}
|k\rangle  \langle k | \sigma | n\rangle + \dots 
\eeq

We may apply this expression to the calculation of the uncertainties over position and momentum:
\beq
\Delta x_n &\approx& \Delta x_n^{(0)} \ 
\left( 1 - \frac{\eta}{2}  \sum_{k\neq n}  \frac{\epsilon_n + \epsilon_k}{\epsilon_n - \epsilon_k}
\frac{\langle k | \sigma | n\rangle}{\left(\Delta x_n^{(0)}\right)^2}  \right. \nonumber \\
&\times& \left. \left( \langle n | x^2 | k \rangle - \langle n | x | n\rangle \langle n | x | k\rangle 
\right) \right) \\
\Delta p_n &\approx& \Delta p_n^{(0)} \ 
\left( 1 - \frac{\eta}{2}  \sum_{k\neq n}  \frac{\epsilon_n + \epsilon_k}{\epsilon_n - \epsilon_k}
\frac{\langle k | \sigma | n\rangle}{\left(\Delta p_n^{(0)}\right)^2}  \right. \nonumber \\
&\times& \left. \left( \langle n | p^2 | k \rangle - \langle n | p | n\rangle \langle n | p | k\rangle 
\right) \right)  \ .
\eeq

We thus obtain the product of the uncertainties
\begin{widetext}
\beq
\Delta x_n \Delta p_n &\approx& \Delta x_n^{(0)} \Delta p_n^{(0)}  \ \left[
 1 - \frac{\eta}{2}  \sum_{k\neq n}  \frac{\epsilon_n + \epsilon_k}{\epsilon_n - \epsilon_k}
\langle k | \sigma | n\rangle \right. \nonumber \\
&\times& \left. \left(
\frac{\left( \langle n | x^2 | k \rangle - \langle n | x | n\rangle \langle n | x | k\rangle 
\right)}{\left(\Delta x_n^{(0)}\right)^2}  +
\frac{\left( \langle n | p^2 | k \rangle - \langle n | p | n\rangle \langle n | p | k\rangle 
\right)}{\left(\Delta p_n^{(0)}\right)^2}  
\right) + O\left[\eta^2\right] \right] \ ,
\eeq
\end{widetext}
which for the simple harmonic oscillator simplifies to
\beq
\Delta x_n \Delta p_n &\approx& \Delta x_n^{(0)} \Delta p_n^{(0)}  \ \left( 1 + O\left[\eta^2\right] \right) \ .
\eeq

\section{Conclusions}
\label{concl} 

Our paper contains strong numerical and analytical results for the solution of the Helmholtz equation
on general domains. We briefly state the novelties and advantages of the approaches outlined in this 
work:
\begin{itemize}
\item The extension of the CMM of Robnik~\cite{Robnik84} to use the orthonormal basis of a square allows to 
      express the integrals (for polynomial conformal mapping or mapping which can be well approximated
      by polynomials) analytically. This clearly reduces the computation times and improves the precision.

\item The computational advantages of the CCM of Amore~\cite{Amore08} are made clear: the collocation matrix 
      is made of two parts, a "universal" matrix, corresponding to the discretization of the negative Laplacian,
      which can be calculated once and for all for a given grid, and a specific "shape dependent" matrix, which
      is diagonal and therefore may be calculated quite efficiently. In this approach no integrals are ever
      needed to be calculated.

\item We prove two theorems which provide explicit formulas for the ground and excited states of a quantum billiard.
      These theorems essentially implement the very well known "power method" for finite dimensional matrices, to
      operators in an infinite dimensional Hilbert space. 
      The variational theorem also allows to obtain upper bounds to the ground state energy. 

\item We have formulated a systematic perturbation theory for membranes obtained slightly deforming a square or
      circular membrane. Our perturbation scheme is essentially the Rayleigh-Schr\"odinger perturbation theory
      with a specific form of the interaction operator. The perturbative results discussed in this paper 
      are in most cases very precise and provide a clear and simple explanation of the mechanism responsible for
      the lifting of degeneracies. For the case of the small deformations of the square, we have obtained
      an analytical formula for the whole spectrum. A possible application of our shape perturbation theory 
      could be to the calculation of Casimir energies due to deformations of membranes. In a recent interesting paper, 
      ref.~\cite{Mazzitelli09}, for example, the point matching method has been used to numerically calculate 
      Casimir energies for perfect-conductor waveguide of arbitrary section. Our perturbative approach could be useful
      for analytical calculations.

\item The shape perturbation theory of the present paper could be applied to study the statistical 
      distribution of the energy level spacings for quantum billiards obtained from small perturbations
      of the square or of the circle. In this case the study would provide an alternative tool to purely numerical
      methods or to asymptotic methods.

\item We have provided a way to extract (at least in principle) the perturbative coefficients of the
      energy of a given level to a specific order. It would be certainly interesting to study the convergence properties
      of the perturbative series corresponding to quantum billiards of different shape and see if a relation 
      between the chaotic behaviour of a billiard and the perturbative series could be established.

\item The analytical results contained in the paper {\sl are not limited} to two dimensional membranes, but 
      can be used for problems of higher dimensions; moreover they can also be applied to describe
      the spectrum of arbitrary inhomogeneous membranes. A numerical study of this problem using the CCM has 
      been already performed in \cite{Amore09}; we plan to look at this problem with the analytical techniques
      of the present paper in the near future.
      Here we have applied our results to Hamiltonians containing position dependent effective masses, extending
      the perturbative approach developed for the Helmholtz equation.

\end{itemize}

We believe that several developments may stem out of the results contained in this paper: among these, it would certainly
be worth looking at possible non-perturbative extensions of our shape perturbation method, which we plan to
look at in the near future. Similarly, we plan to carry out an extension of the perturbative calculation of this paper 
to higher orders. A further interesting application could be to the inverse problem: in a classical paper,
Kac ~\cite{Kac66} posed the question whether one can hear the sound of a drum, meaning by this if the complete 
knowledge of the spectrum of a drum is sufficient to determine its shape. We now know that the answer to this question 
is negative,  since in 1992 Gordon, Webb and Wolpert ~\cite{Gordon92} found an example of two two-dimensional not 
isometric drums which are isospectral. On the other hand, the perturbative formulas obtained in this paper might 
be used to find the shape of a drum  which is obtained from a small deformation of a square or of a circle, knowing a 
number of frequencies. Finally the last application that we wish to mention is to the vibration of drums with 
fractal boundaries: refs.\cite{Lapidus95,Neuberger06, Banjai07}, for example, consider the vibration of a membrane 
whose border is the Koch snowflake. In particular Banjai~\cite{Banjai07} uses a conformal transformation 
to map the snowflake into the unit circle and then performs a spectral collocation to obtain numerical results.
Our perturbative approach, given the conformal map, is directly applicable to this problem and could maybe
provide further insight in this problem.

\begin{acknowledgments}
The author ackowledges support of Conacyt through the SNI program.
\end{acknowledgments}

\appendix

\section{Recurrence relations}
\label{app_a}

We introduce the definitions:
\beq
\phi_{n}(x) &\equiv& \sin \left( \frac{n\pi}{2} (x+1)\right) \\
\chi_{n}(x) &\equiv&  \frac{\delta_{n0} }{\sqrt{2}} + \theta(n-1) \cos \left( \frac{n\pi}{2} (x+1)\right) \ ,
\eeq
where $\theta(n-1)=1$ for $n\geq 1$. 

Notice that these functions obey Dirichlet and von Neumann boundary conditions respectively
and that they are orthonormal:
\beq
\int_{-1}^{+1} \phi_n(x) \phi_m(x) dx  = \int_{-1}^{+1} \chi_n(x) \chi_m(x) dx = \delta_{nm} \ .
\eeq

We consider the integrals:
\beq
\mathcal{Q}_{nmk} &\equiv& \int_{-1}^{+1} dx \ x^k \ \phi_n(x) \phi_m(x) \\
\mathcal{R}_{nmk} &\equiv& \int_{-1}^{+1} dx \ x^k \  \chi_n(x) \chi_m(x) \ .
\eeq

With simple algebra we are able to obtain recurrence relations. First we consider the case $m \neq n$:
\beq
\mathcal{Q}_{nmk}  &=&  - \frac{m^2+n^2}{(k+1)(k+2)} \frac{\pi^2}{4} \mathcal{Q}_{nmk+2} \nonumber \\
&+& \frac{m n}{(k+1)(k+2)} \frac{\pi^2}{2} \mathcal{R}_{nmk+2} \\
\mathcal{R}_{nmk}  &=&  \frac{(-1)^k+(-1)^{m+n}}{k+1}- \frac{m^2+n^2}{(k+1)(k+2)} \frac{\pi^2}{4} \mathcal{R}_{nmk+2} \nonumber \\
&+& \frac{m n}{(k+1)(k+2)} \frac{\pi^2}{2} \mathcal{Q}_{nmk+2} \ .
\eeq

These equations may be solved to yield the recurrence relations:
\beq
\mathcal{Q}_{nmk+2} &=& \frac{8 (k+2) m n \left((-1)^k+(-1)^{m+n}\right)}{\pi^2 \left(m^2-n^2\right)^2} \nonumber \\
&-&\frac{4 (k+1) (k+2) \left(m^2+n^2\right)}{\pi^2 \left(m^2-n^2\right)^2} \mathcal{Q}_{nmk} \nonumber \\
&-& \frac{8 (k+1) (k+2) m n}{\pi^2 \left(m^2-n^2\right)^2} \mathcal{R}_{nmk}\\
\mathcal{R}_{nmk+2} &=& \frac{4 (k+2) \left(m^2+n^2\right) \left((-1)^k+(-1)^{m+n}\right)}{\pi^2 \left(m^2-n^2\right)^2} \nonumber \\
&-& \frac{8 (k+1) (k+2) m n}{\pi^2 \left(m^2-n^2\right)^2} \mathcal{Q}_{nmk} \nonumber \\
&-&\frac{4 (k+1) (k+2) \left(m^2+n^2\right)}{\pi^2 \left(m^2-n^2\right)^2}\mathcal{R}_{nmk}
\eeq

For $m=n$ we have
\beq
\mathcal{Q}_{nmk} + \mathcal{R}_{nmk} = \frac{1+(-1)^k}{k+1}
\eeq
and
\beq
\mathcal{Q}_{nmk}  &=& \frac{\pi^2 \left((-1)^k+1\right) n^2}{2 \left(k^3+6 k^2+11 k+6\right)} \nonumber \\
&-&\frac{\pi^2 n^2}{k^2+3 k+2} \mathcal{Q}_{nmk+2} 
\eeq
which is easily solved to give
\beq
\mathcal{Q}_{nmk+2}  &=& \frac{(-1)^k+1}{2 (k+3)}-\frac{k^2+3 k+2}{\pi^2 n^2} \ \mathcal{Q}_{nmk} \ .
\eeq

After calculating explicitly $\mathcal{Q}_{nm0}$, $\mathcal{Q}_{nm1}$, $\mathcal{R}_{nm0}$ and $\mathcal{R}_{nm1}$ 
one may then obtain the coefficients of higher order by simply applying the recurrence relations.

For instance:
\beq
\mathcal{Q}_{nm0} &=& \delta_{nm} \nonumber\\
\mathcal{Q}_{nn1} &=& 0 \ , \ m = n \nonumber\\
\mathcal{Q}_{nm1} &=& \frac{8 m n \left((-1)^{m+n}-1\right)}{\pi^2 \left(m^2-n^2\right)^2} \ , \ m \neq n \nonumber\\
\mathcal{Q}_{nn2} &=& \frac{1}{3}-\frac{2}{\pi^2 n^2} \ , \ m = n \nonumber\\
\mathcal{Q}_{nm2} &=& \frac{16 m n \left((-1)^{m+n}+1\right)}{\pi^2 \left(m^2-n^2\right)^2}  \ , \ m = n \nonumber\\
\mathcal{Q}_{nn3} &=& 0 \ , \ m = n \nonumber\\
\mathcal{Q}_{nm3} &=&  \frac{24 m n \left((-1)^{m+n}-1\right) }{\pi^4 \left(m^2-n^2\right)^4} \nonumber\\
&\times& \left(\pi^2 \left(m^2-n^2\right)^2-16 \left(m^2+n^2\right)\right) 
\ , \ m \neq n \nonumber \\
&\dots& \nonumber 
\eeq

\section{Perturbation theory}
\label{app_b}

In this section we give some detail on the calculation of the perturbative coefficients
for the energy. In the following we introduce the definition $\omega_{nk} \equiv \epsilon_n -
\epsilon_k$. We may use the expressions given in eqns.(\ref{pt4}),(\ref{pt5}),(\ref{pt6})
and (\ref{pt7}) inside the general formulas of eqns.(\ref{pt8}),(\ref{pt9}),(\ref{pt10}) 
and (\ref{pt11}) to obtain the analytical expressions which are reported in the paper.

We start with the second order term, since the zero and first order terms are straightforward.
In this case we obtain:
\beq
E_n^{(2)} &=& \frac{1}{4} \left[ \sum_l \langle n | \sigma | l \rangle \epsilon_l 
+ 3 \epsilon_n \langle n | \sigma^2 | n \rangle  \right]  \nonumber \\
&+& \frac{1}{4} \sum_{k \neq n} \frac{\langle n | \sigma | k \rangle^2 }{\omega_{nk}} (\epsilon_n+\epsilon_k)^2
\eeq

Notice that the first sum in this expression is unrestricted, since $l$ can take all possible value,
including $l=n$. It is convenient to express the unrestricted sums isolating the term corresponding to
$l=n$, i.e. writing $\sum_l f(l) = f(n) + \sum_{l\neq n}f(l)$. 

Keeping in mind this observation we may rewrite the first term of the equation above as
\beq
\left. E_n^{(2)}\right|_I  &=& \frac{1}{4} \left[ \sum_l \langle n | \sigma | l \rangle \epsilon_l 
+ 3 \epsilon_n \langle n | \sigma^2 | n \rangle  \right]  \nonumber \\
&=& \frac{3}{4}\epsilon_n \langle n | \sigma^2 | n \rangle + \frac{1}{4} \epsilon_n 
\langle n | \sigma | n \rangle^2 + \frac{1}{4} \sum_{k\neq n} \langle n | \sigma | k \rangle^2  \epsilon_{k} 
\nonumber \ .
\eeq

Similarly we may simplify the second term:
\beq
\left. E_n^{(2)}\right|_{II}  &=& 
\frac{1}{4} \sum_{k \neq n} \frac{\langle n | \sigma | k \rangle^2 }{\omega_{nk}} (\epsilon_n+\epsilon_k)^2 \nonumber \\
&=&  \epsilon_n^2 \sum_{k \neq n}  \frac{\langle n | \sigma | k \rangle^2 }{\omega_{nk}} \nonumber \\
&-&  \epsilon_n \sum_{k\neq n} \langle n | \sigma | k \rangle^2 + \frac{1}{4} \sum_{k\neq n} 
\langle n | \sigma | k \rangle^2  \omega_{nk} \nonumber \\
&=& \epsilon_n^2 \sum_{k \neq n}  \frac{\langle n | \sigma | k \rangle^2 }{\omega_{nk}} 
- \epsilon_n \left(\langle n | \sigma^2 | n \rangle - \langle n | \sigma | n \rangle^2 \right) \nonumber \\
&+& \frac{\epsilon_n}{4} \left(\langle n | \sigma^2 | n \rangle - \langle n | \sigma | n \rangle^2 \right)
- \frac{1}{4} \sum_{k\neq n} \langle n | \sigma | k \rangle^2  \epsilon_{k}  \nonumber \ .
\eeq

Combining the two terms we obtain the final result:
\beq
E_n^{(2)} = \epsilon_n \langle n | \sigma | n \rangle^2 
+ \epsilon_n^2 \sum_{k \neq n} \frac{\langle n | \sigma | k \rangle^2 }{\epsilon_n-\epsilon_k} \ .
\eeq

\begin{widetext}
The third order contribution reads:

\beq
E_n^{(3)} &=& \frac{3}{8} \sum_l \omega_{nl} \langle n | \sigma^2 | l \rangle \langle l | \sigma | n \rangle 
- \epsilon_n \langle n | \sigma^3 | n \rangle 
- \frac{3}{8} \sum_{k\neq n} \langle n | \sigma | k \rangle \langle k | \sigma^2 | n \rangle
\frac{(\epsilon_n+\epsilon_k)^2}{\omega_{nk}} \nonumber \\
&-& \frac{1}{4}  \sum_{k\neq n} \sum_l \langle n | \sigma | k \rangle \langle k | \sigma | l \rangle
\langle l | \sigma | n \rangle \frac{\epsilon_l (\epsilon_n+\epsilon_k)}{\omega_{nk}} 
- \frac{1}{8}  \sum_{k\neq n} \sum_{m \neq n} \langle n | \sigma | k \rangle \langle k | \sigma | m \rangle
\langle m | \sigma | n \rangle 
\frac{(\epsilon_n+\epsilon_k)  (\epsilon_n+\epsilon_m)  (\epsilon_m+\epsilon_k)}{\omega_{nk} \omega_{nm}} \nonumber \\
&+& \frac{1}{4} \epsilon_n \langle n | \sigma | n \rangle 
\sum_{k\neq n} \langle n | \sigma | k \rangle^2 \frac{(\epsilon_n+\epsilon_k)^2}{\omega_{nk}^2}  \ .
\label{appB1}
\eeq

We may simplify this expression by arranging term with equal number of sums together. 
Keeping in mind this simple observation we consider the terms in the above expression which contain two sums:

\beq
\left. E_n^{(3)}\right|_{2 sums} &=& - \frac{1}{8}  \sum_{k\neq n} \sum_{m \neq n} \langle n | \sigma | k \rangle \langle k | \sigma | m \rangle
\langle m | \sigma | n \rangle \left[ \frac{\epsilon_m (\epsilon_n+\epsilon_k)}{\omega_{nk}} +
\frac{\epsilon_k (\epsilon_n+\epsilon_m)}{\omega_{nm}} +
\frac{(\epsilon_n+\epsilon_k)  (\epsilon_n+\epsilon_m)  (\epsilon_m+\epsilon_k)}{\omega_{nk} \omega_{nm}}\right]  
\nonumber \\
&=&  -\frac{\epsilon_n}{4} \sum_{k\neq n} \sum_{m \neq n} \langle n | \sigma | k \rangle \langle k | \sigma | m \rangle
\langle m | \sigma | n \rangle \left[ 2 - 3 \epsilon_n \left(\frac{1}{\omega_{nk}}+\frac{1}{\omega_{nm}} \right)
+ 4 \frac{\epsilon_n^2}{\omega_{nk}\omega_{nm}} \right] \ .
\label{appB2}
\eeq

We start with the first term in the square parenthesis:
\beq
\left. E_n^{(3)}\right|_{2 sums}^{I} &=& - \frac{\epsilon_n}{2} \left( \langle n | \sigma^3 | n \rangle 
- 2 \langle n | \sigma^2 | n \rangle \langle n | \sigma | n \rangle   + \langle n | \sigma | n \rangle^3
\right) \ .
\label{appB3}
\eeq

We now come to the second and term terms in the square parethesis: a simple relabeling of the summed indices 
shows us that these terms are equal and therefore:
\beq
\left. E_n^{(3)}\right|_{2 sums}^{II} &=& \frac{3}{2} \epsilon_n^2 \sum_{k\neq n} \sum_{m \neq n} 
\frac{\langle n | \sigma | k \rangle \langle k | \sigma | m \rangle
\langle m | \sigma | n \rangle }{\omega_{nk}} \nonumber \\
&=& \frac{3}{2} \epsilon_n^2 \sum_{k\neq n}  \frac{\langle n | \sigma | k \rangle }{\omega_{nk}} 
\left[\langle k | \sigma^2 | n \rangle -\langle k | \sigma | n \rangle \langle n | \sigma | n \rangle
\right] \ .
\label{appB4}
\eeq
Finally we write the last term as:
\beq
\left. E_n^{(3)}\right|_{2 sums}^{III} &=& 
-\epsilon_n^3 \sum_{k\neq n} \sum_{m \neq n} 
\frac{\langle n | \sigma | k \rangle \langle k | \sigma | m \rangle 
\langle m | \sigma | n \rangle }{\omega_{nk}\omega_{nm}} 
\label{appB5}
\eeq

We may now write
\beq
E_n^{(3)} &=& \frac{3}{8} \sum_{l\neq n} \omega_{nl} \langle n | \sigma^2 | l \rangle \langle l | \sigma | n \rangle 
- \epsilon_n \langle n | \sigma^3 | n \rangle 
- \frac{3}{8} \sum_{k\neq n}  \langle n | \sigma | k \rangle \langle k | \sigma^2 | n \rangle
\frac{(\epsilon_n+\epsilon_k)^2}{\omega_{nk}} \nonumber \\
&-& \frac{1}{4}  \sum_{k\neq n} \langle n | \sigma | k \rangle \langle k | \sigma | n \rangle
\langle n | \sigma | n \rangle \frac{\epsilon_n (\epsilon_n+\epsilon_k)}{\omega_{nk}} 
+ \frac{1}{4} \epsilon_n \langle n | \sigma | n \rangle 
\sum_{k\neq n} \langle n | \sigma | k \rangle^2 \frac{(\epsilon_n+\epsilon_k)^2}{\omega_{nk}^2}  \nonumber \\
&-& \frac{\epsilon_n}{2} \left( \langle n | \sigma^3 | n \rangle 
- 2 \langle n | \sigma^2 | n \rangle \langle n | \sigma | n \rangle   + \langle n | \sigma | n \rangle^3
\right) + \frac{3}{2} \epsilon_n^2 \sum_{k\neq n}  \frac{\langle n | \sigma | k \rangle }{\omega_{nk}} 
\left[\langle k | \sigma^2 | n \rangle -\langle k | \sigma | n \rangle \langle n | \sigma | n \rangle
\right] \nonumber \\
&-& \epsilon_n^3 \sum_{k\neq n} \sum_{m \neq n} 
\frac{\langle n | \sigma | k \rangle \langle k | \sigma | m \rangle 
\langle m | \sigma | n \rangle }{\omega_{nk}\omega_{nm}}  \ .
\label{appB6}
\eeq

With simple algebra we obtain:
\beq
E_n^{(3)} &=& - \epsilon_n \langle n | \sigma | n \rangle^3 + \epsilon_n^3 \langle n | \sigma | n \rangle
\sum_{k \neq n}  \frac{\langle n | \sigma | k \rangle^2}{\omega_{nk}^2}
- 3 \epsilon_n^2  \langle n | \sigma | n \rangle \sum_{k \neq n}  \frac{\langle n | \sigma | k \rangle^2}{\omega_{nk}}
\nonumber \\
&-& \epsilon_n^3 \sum_{k\neq n} \sum_{m \neq n} 
\frac{\langle n | \sigma | k \rangle \langle k | \sigma | m \rangle 
\langle m | \sigma | n \rangle }{\omega_{nk}\omega_{nm}}  \ .
\label{appB7}
\eeq

\end{widetext}

\begin{table}[!htb]
\begin{tabular}{|c||c|c|c|}
	\hline
$N$	&  $E_0$    &   $E_{1,2}$  & $E_{3,4}$   \\
	\hline \hline
$10$   & 5.7831903186 & 14.682015349 & 26.374703996 \\
$20$   & 5.7831860077 & 14.681971199 & 26.374617574 \\
$30$   & 5.7831859657 & 14.681970679 & 26.374616505 \\  
$40$   & 5.7831859633 & 14.681970647 & 26.374616438 \\ 
$50$   & 5.7831859630 & 14.681970643 & 26.374616430  \\ 
$60$   & 5.7831859630 & 14.681970642 & 26.374616428  \\
	\hline \hline
exact  & 5.7831859629 & 14.681970642 & 26.374616427 \\
	\hline
\end{tabular}
\caption{Three lowest energy eigenvalues of a circular membrane of unit radius obtained
using the Collocation Mapping Method with different number of elements. 
The conformal map is approximated with the Taylor series up to order $z^{37}$.}
\label{table-1}
\end{table}

\begin{table}[!htb]
\begin{tabular}{|c||c|c|c|}
	\hline
$N$	&  $E_0$    &   $E_{1,2}$  & $E_{3,4}$   \\
	\hline \hline
$20$   & 5.7833478471 & 14.683030231 & 26.376321659 \\
$40$   & 5.7831962133 & 14.682036989 & 26.374723431 \\
$60$   & 5.7831879924 & 14.681983747 & 26.374637569 \\
$80$   & 5.7831866056 & 14.681974789 & 26.374623117 \\
$100$  & 5.7831862262 & 14.681972340 & 26.374619167 \\
	\hline \hline
exact  & 5.7831859629 & 14.681970642 & 26.374616427 \\
	\hline
\end{tabular}
\caption{Three lowest energy eigenvalues of a circular membrane of unit radius obtained
using the Collocation Collocation Method with different grid size.}
\label{table-2}
\end{table}

\begin{table}[htb]
\begin{tabular}{|c|c|c|c|c|c|c|}
	\hline
$(n_x,n_y)$ & $d$ &PT$_0$	&  PT$_1$    &   PT$_2$  &  PT$_3$  & exact \\
	\hline \hline
$(1,1)$ & 1 &  4.93480 & 5.66472   &  5.76740  &   5.78118  & 5.78319  \\
\hline
$(1,2)$ & 2 &  12.3370 & 14.4197   &  14.6695  &   14.6844  & 14.68197 \\
$(2,1)$ &   &  12.3370 & 14.4197   &  14.6695  &   14.6844  & 14.68197 \\
\hline
$(2,2)$ & 1 &  19.7392 & 25.0228   &  26.1714  &   26.3593  & 26.37462 \\
\hline
$(1,3)$ & 2 &  24.6740 & 26.4137   &  26.4007  &   26.3785  & 26.37462 \\
$(3,1)$ &   &  24.6740 & 30.2612   &  30.7573  &   30.5467  & 30.47126 \\
\hline
$(2,3)$ & 2 &  32.0762 & 40.3174   &  41.4936  &   41.1156  & 40.70647 \\
$(3,2)$ &   &  32.0762 & 40.3174   &  41.4936  &   41.1156  & 40.70647 \\
\hline 
$(1,4)$ & 2 &  41.9458 & 47.7922   &  48.5522  &   48.9791  & 49.21846 \\
$(4,1)$ &   &  41.9458 & 47.7922   &  48.5522  &   48.9791  & 49.21846 \\
\hline 
$(3,3)$ & 1 &  44.4132 & 55.1980   &  57.6642  &   57.9679  & 57.58294 \\
\hline
$(2,4)$ & 2 &  49.3480 & 57.1883   &  57.7144  &   57.6217  & 57.58294 \\
$(4,2)$ &   &  49.3480 & 66.4508   &  71.4408  &   72.0826  & 70.85000 \\
\hline
$(3,4)$ & 2 &  61.6850 & 76.3544   &  77.2464  &   76.2107 &  76.93893 \\
$(4,3)$ &   &  61.6850 & 76.3544   &  77.2464  &   76.2107 &  76.93893 \\
\hline
$(1,5)$ & 2 &  64.1524 & 72.9892   &  71.1717  &   70.3712 &  70.85000 \\
$(5,1)$ &   &  64.1524 & 72.5674   &  73.5746  &   74.6076 &  74.88701 \\
\hline
$(5,2)$ & 2 &  71.5546 & 89.4650   &  96.8822  &   99.3590 &  95.27757\\
$(2,5)$ &   &  71.5546 & 89.4650   &  96.8822  &   99.3590 &  95.27757\\
\hline
$(4,4)$ & 1 &  78.9568 & 97.3221   &  97.4219  &   96.4281  & 98.72627 \\
	\hline 
\end{tabular}
\caption{Energy of the twenty lowest states of the circular membrane calculated with
perturbation theory to different orders. }
\label{table-3}
\end{table}

\begin{widetext}

\begin{table}[htb]
\begin{tabular}{|r|r|r|r|r|r|r|c|}
	\hline
$n$ & $\alpha=\frac{1}{25}$ & $\alpha=\frac{1}{50}$	&  $\alpha = \frac{1}{100}$  &  $\alpha = \frac{1}{200}$ &  $\alpha = \frac{1}{400}$ &  $\alpha = \frac{1}{800}$
& PT  \\
\hline \hline
1   &   -11.9314  &   -11.9828  &   -11.9957  &     -11.9989  &   -11.9997  &   -11.9999  &  -12          \\
2   &   -62.2270  &   -62.9170  &   -63.0930  &     -63.1372  &   -63.1482  &   -63.1510  &  -63.15197799 \\
3   &    -7.74588 &    -7.6738  &    -7.6556  &      -7.6510  &    -7.6498  &    -7.6495  &  -7.649505501 \\
4   &   -12.8722  &   -12.2228  &   -12.0560  &     -12.0139  &   -12.0034  &   -12.0007  &  -12          \\
5   &  -207.2669  &  -212.6755  &  -214.1045  &    -214.4669  &  -214.5578  &  -214.5805  &  -214.5883271 \\
6   &   -13.27723 &   -13.2145  &   -13.1989  &     -13.1950  &   -13.1940  &   -13.1938  &  -13.19388958 \\
7   &   -25.0501  &   -24.8644  &   -24.8184  &     -24.8069  &   -24.8040  &   -24.8033  &  -24.80336862 \\
8   &     3.2132  &     8.2394  &     9.5815  &       9.9228  &    10.0085  &    10.0299  &   10.03674606 \\
9   &  -529.3273  &  -558.6905  &  -567.0588  &    -569.2278  &  -569.7751  &  -569.9123  &  -569.9586259 \\
10  &   -22.0706  &   -21.9810  &   -21.9587  &     -21.9531  &   -21.9517  &   -21.9513  &  -21.95180463 \\
11  &   -12.8590  &   -12.2085  &   -12.0512  &     -12.0123  &   -12.0026  &   -12.0002  &  -12          \\
12  &   -44.8505  &   -44.7072  &   -44.6716  &     -44.6627  &   -44.6605  &   -44.6599  &  -44.66052201 \\
13  &    46.7717  &    74.1811  &    82.1277  &      84.1959  &    84.7183  &    84.8492  &   84.89208802 \\
14  &   -39.7638  &   -39.5706  &   -39.5237  &     -39.5120  &   -39.5091  &   -39.5084  &  -39.50940841 \\
15  &    30.6629  &    33.2746  &    33.8759  &      34.0227  &    34.0591  &    34.0682  &   34.0699668  \\
16  & -1109.8341  & -1223.2138  & -1259.7952  &   -1269.6749  & -1272.1960  & -1272.8294  &  -1273.042285 \\
17  &   -33.6433  &   -33.4920  &   -33.4543  &     -33.4449  &   -33.4425  &   -33.4419  &  -33.44310862\\
18  &   -70.2925  &   -70.1087  &   -70.0628  &     -70.0513  &   -70.0484  &   -70.0477  &  -70.04918774 \\
19  &   102.5451  &   207.3338  &   242.1806  &     251.6572  &   254.0794  &   254.6883  &   254.8899234 \\
20  &   -12.5604  &   -12.1285  &   -12.0299  &     -12.0059  &   -11.9999  &   -11.9984  &  -12\\
21  &   -70.6796  &   -70.5877  &   -70.5651  &     -70.5595  &   -70.5581  &   -70.5577  &  -70.55997503\\
22  &   124.8295  &   134.9087  &   137.0858  &     137.6039  &   137.7317  &   137.7635  &   137.7717825 \\
23  & -2007.8189  & -2335.7052  & -2460.1854  &   -2496.0792  & -2505.4160  & -2507.7742  &  -2508.565211 \\
24  &   -47.9136  &   -47.6534  &   -47.5887  &     -47.5725  &   -47.5685  &   -47.5675  &  -47.56994783\\
25  &  -101.2945  &  -101.0139  &  -100.9438  &    -100.9262  &  -100.9219  &  -100.9208  &  -100.9237065\\
26  &   115.2123  &   410.2533  &   528.4606  &     562.9517  &   571.9490  &   574.2230  &   574.980025 \\
27  &   -53.7499  &   -53.6334  &   -53.6067  &     -53.6002  &   -53.5986  &   -53.5981  &  -53.60145365\\
28  &    50.1431  &    51.1784  &    51.3848  &      51.4331  &    51.4450  &    51.4480  &   51.44555258\\
29  &  -106.3989  &  -106.3559  &  -106.3450  &    -106.3423  &  -106.3416  &  -106.3414  &  -106.3455495\\
30  &   286.6324  &   322.1690  &   329.4268  &     331.0946  &   331.5016  &   331.6027   &  331.632198 \\
31  & -3256.1580  & -4011.4680  & -4354.9098  &   -4463.3391  & -4492.3946  & -4499.7924   &  -4502.275663 \\
32  &   -64.8743  &   -64.4402  &   -64.3324  &     -64.3055  &   -64.2988  &   -64.2971   &  -64.30174616\\
33  &   -12.0909  &   -12.0078  &   -11.9974  &     -11.9955  &   -11.9950  &   -11.9949   &  -12\\
34  &   -96.1084  &   -96.0849  &   -96.0799  &     -96.0787  &   -96.0784  &   -96.0783   &  -96.08384485\\
35  &   167.0112  &   169.8399  &   170.2940  &     170.3920  &   170.4156  &   170.4214   &  170.4178176\\
36  &  -137.8993  &  -137.4572  &  -137.3467  &    -137.3190  &  -137.3121  &  -137.3104   &  -137.3156766\\
37  &    -0.2977  &   647.4859  &   970.7556  &    1074.8721  &  1102.9016  &  1110.0463   &   1112.43464 \\
38  &  -147.5040  &  -147.5073  &  -147.5074  &    -147.5074  &  -147.5073  &  -147.5073   &   -147.5143637\\
39  &   516.9942  &   627.1522  &   649.2405  &     654.1087  &   655.2782  &   655.5675   &   655.6568115 \\
40  &   -67.3226  &   -67.3763  &   -67.3928  &     -67.3972  &   -67.3982  &   -67.3985   &  -67.40630759\\
41  &    67.0772  &    66.9190  &    66.8387  &      66.8163  &    66.8105  &    66.8091   &   66.8009496\\
42  & -4871.0690  & -6337.4062  & -7134.0401  &   -7415.2895  & -7493.8895  & -7514.1430   &   -7520.958743\\ 
43  &  -141.9693  &  -142.0124  &  -142.0235  &    -142.0262  &  -142.0270  &  -142.0271   &   -142.0360122\\
44  &   -84.5423  &   -83.8471  &   -83.6747  &     -83.6317  &   -83.6210  &   -83.6183   &   -83.62637707\\
45  &   359.2556  &   367.9325  &   369.0389  &     369.2503  &   369.2992  &   369.3111   &   369.3064766\\
46  &  -330.0113  &  -179.4676  &  -179.2960  &    -179.2531  &  -179.2424  &  -179.2397   &   -179.2486248\\
47  &  -180.1538  &   837.1280  &  1575.4900  &    1844.8179  &  1920.6443  &  1940.2181   &   1946.790496\\
48  &   -11.4814  &   -11.8521  &   -11.9544  &     -11.9805  &   -11.9871  &   -11.9887   &   -12\\
49  &  -194.2969  &  -194.3597  &  -194.3737  &    -194.3770  &  -194.3779  &  -194.3781   &  -194.3894759\\
50  &   785.2942  &  1078.8172  &  1140.1836  &    1153.1570  &  1156.2105  &  1156.9613  &  1157.199583 \\
\hline
\end{tabular}
\caption{Corrections to the energy of a particle in a box of square $2$ for deformations of the square,
$(E_n - E_n^{box})/\alpha^2$, for different values of $\alpha$ for the first 50 energy levels. 
In the last column  are the results of Perturbation Theory.} 
\label{table-pt}
\end{table}

\end{widetext}

\begin{widetext}

\begin{table}[htb]
\begin{tabular}{|c|c|}
	\hline
$(n_x,n_y)$ & $-E_n^{(1)}/E_n^{box}$ \\
	\hline \hline
$(1,1)$ & $2\varrho_1 + \left(\frac{48}{\pi^4}-\frac{8}{15}\right) \varrho_5+\left(\frac{32}{45}+\frac{80640}{\pi^8}
        - \frac{896}{\pi^4}\right) \varrho_9
        + \left(-\frac{128}{91}+\frac{958003200}{\pi^{12}}-\frac{10644480}{\pi^8}+\frac{8448}{\pi^4}\right) \varrho_{13}$ \\
        & $+\left(\frac{512}{153}+\frac{41845579776000}{\pi^{16}}-\frac{464950886400}{\pi^{12}}+\frac{369008640}{\pi^8}
        - \frac{61440}{\pi^4}\right) \varrho_{17} + \dots$  \\
\hline
$(1,2)$ & $2\varrho_1 -\frac{3 \varrho_3}{\pi^2}+\left(\frac{39}{\pi^4}-\frac{8}{15}\right) \varrho_5+
          \frac{3 \left(8 \pi^4-765\right) \varrho_7}{2 \pi^6}+\left(\frac{32}{45}+\frac{64575}{\pi^8}
        - \frac{728}{\pi^4}\right) \varrho_9$ \\
$(2,1)$ & $-\frac{3 \left(3869775-42840 \pi^4+32 \pi^8\right) \varrho_{11}}{2 \pi^{10}}
        + \left(-\frac{128}{91}+\frac{1532898675}{2 \pi^{12}}-\frac{8523900}{\pi^8}+\frac{6864}{\pi^4}\right) \varrho_{13}$ \\
        & $+\frac{3 \left(-185977516725+2065943880 \pi^4-1633632 \pi^8+256 \pi^{12}\right) \varrho_{15}}{4 \pi^{14}} 
        + \left(\frac{512}{153}+\frac{33476591523375}{\pi^{16}}-\frac{371983411800}{\pi^{12}}+\frac{295495200}{\pi^8}
        - \frac{49920}{\pi^4}\right) \varrho_{17}$ \\
        & $-\frac{3 \left(6829192106611875-75878826823800 \pi^4+60207507360 \pi^8-9987840 \pi^{12}
        + 512 \pi^{16}\right) \varrho_{19}}{2 \pi^{18}} + \dots$  \\
\hline
$(2,2)$ & $2\varrho_1  + \left(\frac{3}{\pi^4}-\frac{8}{15}\right) \varrho_5+\left(\frac{32}{45}+\frac{315}{\pi^8}
        - \frac{56}{\pi^4}\right) \varrho_9+\left(-\frac{128}{91}+\frac{467775}{2 \pi^{12}}-\frac{41580}{\pi^8}
        + \frac{528}{\pi^4}\right) \varrho_{13}$ \\
        & $+\left(\frac{512}{153}+\frac{638512875}{\pi^{16}}-\frac{113513400}{\pi^{12}}+\frac{1441440}{\pi^8}
        - \frac{3840}{\pi^4}\right) \varrho_{17} + \dots$   \\
\hline
$(1,3)^\dagger$ & $\left[2 \varrho_1-\frac{32 \varrho_3}{9 \pi^2}+\frac{8}{135} \left(\frac{730}{\pi^4}-9\right) \varrho_5
        + \frac{128\left(9 \pi^4-820\right) \varrho_7}{81 \pi^6}+\frac{32 \left(8267000-91980 \pi^4+81 \pi^8\right) 
          \varrho_9}{3645 \pi^8} +\dots \right]$ \\
$(3,1)$ & ${\bf \pm} \left[ \frac{27 \varrho_5}{\pi^4}-\frac{63 \left(8 \pi^4-765\right) \varrho_9}{\pi^8} +\dots\right]$ \\
\hline
$(2,3)$ & $2\varrho_1  -\frac{5 \varrho_3}{9 \pi^2}+\left(\frac{61}{27 \pi^4}-\frac{8}{15}\right) \varrho_5
        + \frac{5 \left(72 \pi^4-485\right) 
          \varrho_7}{162 \pi^6}+\left(\frac{32}{45}+\frac{161735}{729 \pi^8}-\frac{3416}{81 \pi^4}\right) \varrho_9$ \\
$(3,2)$ & $-\frac{5 \left(1419775-244440 \pi^4+2592 \pi^8\right) \varrho_{11}}{1458 \pi^{10}}+\left(-\frac{128}{91}
        + \frac{710673425}{4374 \pi^{12}}-\frac{7116340}{243 \pi^8}+\frac{10736}{27 \pi^4}\right) \varrho_{13}$ \\
        & $+\frac{5 \left(-115834293575+20465204760 \pi^4-251675424 \pi^8+559872 \pi^{12}\right) \varrho_{15}}{78732 \pi^{14}}
        + \left(\frac{512}{153}+\frac{26120117398375}{59049 \pi^{16}}-\frac{517370253400}{6561 \pi^{12}}
        + \frac{740099360}{729 \pi^8}-\frac{78080}{27 \pi^4}\right) \varrho_{17}$ \\
       & $-\frac{5 \left(798494934111875-141781175335800 \pi^4+1789243616160 \pi^8
        - 4616144640 \pi^{12}+3359232 \pi^{16}\right) \varrho_{19}}{118098 \pi^{18}}+ \dots$   \\
\hline 
$(1,4)$ & $2\varrho_1  -\frac{15 \varrho_3}{4 \pi^2}+\left(\frac{723}{16 \pi^4}-\frac{8}{15}\right) \varrho_5
        + \left(\frac{15}{\pi^2}-\frac{173475}{128 \pi^6}\right) \varrho_7 +\left(\frac{32}{45}+\frac{19429515}{256 \pi^8}
        - \frac{1687}{2 \pi^4}\right) \varrho_9$ \\
$(4,1)$ & $-\frac{15 \left(932615775-10362240 \pi^4+8192 \pi^8\right) \varrho_{11}}{2048 \pi^{10}}
        + \left(-\frac{128}{91}+\frac{7386317405775}{8192 \pi^{12}}-\frac{641173995}{64 \pi^8}+\frac{7953}{\pi^4}\right) \varrho_{13}$ \\
        & $+\frac{15 \left(-716965206682725+7966279601280 \pi^4-6322348032 \pi^8+1048576 \pi^{12}\right) \varrho_{15}}{65536 \pi^{14}}$ \\
        & $+\left(\frac{512}{153}+\frac{2581074744696322875}{65536 \pi^{16}}-\frac{224051627975175}{512 \pi^{12}}
        + \frac{2778420645}{8 \pi^8}-\frac{57840}{\pi^4}\right) \varrho_{17}$ \\
        & $+\left(-\frac{6318470974918905928125}{524288 \pi^{18}}+\frac{548478383112284625}{4096 \pi^{14}}
        - \frac{6801566847075}{64 \pi^{10}}+\frac{17694450}{\pi^6}-\frac{960}{\pi^2}\right) \varrho_{19}+ \dots$   \\
\hline 
$(3,3)$ & $2\varrho_1 + \left(\frac{16}{27 \pi^4}-\frac{8}{15}\right) \varrho_5+\frac{32 \left(81+\frac{1400}{\pi^8}
        - \frac{1260}{\pi^4}\right) \varrho_9}{3645}+\left(-\frac{128}{91}+\frac{3942400}{2187 \pi^{12}}-\frac{394240}{243 \pi^8}
         + \frac{2816}{27 \pi^4}\right) \varrho_{13}$ \\
        & $+\frac{512 \left(1905904000-1715313600 \pi^4+110270160 \pi^8-1487160 \pi^{12}
        + 6561 \pi^{16}\right) \varrho_{17}}{1003833 \pi^{16}} + \dots$   \\
\hline
$(2,4)^\dagger$ & $\left[ 2 \varrho_1-\frac{3 \varrho_3}{4 \pi^2}+\left(\frac{39}{16 \pi^4}-\frac{8}{15}\right) 
        \varrho_5+\left(\frac{3}{\pi^2}-\frac{2295}{128 \pi^6}\right) \varrho_7+\left(\frac{32}{45}+\frac{64575}{256 \pi^8}
        -\frac{91}{2 \pi^4}\right) \varrho_9 + \dots \right]$ \\
$(4,2)$ &  ${\bf \pm} \left[ \frac{1024 \varrho_5}{27 \pi^4}-\frac{57344 \left(9 \pi^4-820\right) \varrho_9}{729 \pi^8} + \dots \right]$\\
\hline
$(3,4)$ & $2\varrho_1 -\frac{7 \varrho_3}{36 \pi^2}+\left(\frac{193}{432 \pi^4}-\frac{8}{15}\right) \varrho_5
        + \frac{7 \left(1152 \pi^4-1685\right) \varrho_7}{10368 \pi^6}+\left(\frac{32}{45}+\frac{1550675}{186624 \pi^8}
        - \frac{1351}{162 \pi^4}\right) \varrho_9$ \\
$(4,3)$ & $-\frac{7 \left(16245775-13587840 \pi^4+663552 \pi^8\right) \varrho_{11}}{1492992 \pi^{10}} 
        + \left(-\frac{128}{91}+\frac{21037818725}{17915904 \pi^{12}}-\frac{17057425}{15552 \pi^8}
        + \frac{2123}{27 \pi^4}\right) \varrho_{13}$ \\
        & $+\frac{7 \left(-4256172495575+3746769586560 \pi^4-223840641024 \pi^8
        + 2293235712 \pi^{12}\right) \varrho_{15}}{1289945088 \pi^{14}}$ \\
        & $+\left(\frac{512}{153}+\frac{2421160144277875}{3869835264 \pi^{16}}-\frac{1914441503975}{3359232 \pi^{12}}
        +\frac{221746525}{5832 \pi^8}-\frac{15440}{27\pi^4}\right) \varrho_{17}$ \\
        & $-\frac{7 \left(93255273798561875-83352882153340800 \pi^4+5241195398799360 \pi^8-65689736970240 \pi^{12}
        + 220150628352 \pi^{16}\right) \varrho_{19}}{30958682112 \pi^{18}} + \dots$   \\
\hline
$(1,5)^\dagger$ &  $\left[ 2 \varrho_1-\frac{96 \varrho_3}{25 \pi^2}+\left(\frac{28848}{625 \pi^4}-\frac{8}{15}\right) \varrho_5+\frac{384
          \left(125 \pi^4-11268\right) \varrho_7}{3125 \pi^6}+\left(\frac{32}{45}+\frac{6057692928}{78125 \pi^8}
        - \frac{538496}{625 \pi^4}\right) \varrho_9 + \dots \right]$ \\
$(5,1)$ & ${\bf \pm} \left[ \frac{25 \varrho_5}{27 \pi^4}-\frac{175 \left(72 \pi^4-485\right) \varrho_9}{729 \pi^8} +\dots\right]$\\
\hline
$(5,2)$ &  $2\varrho_1-\frac{21 \varrho_3}{25 \pi^2}+\left(\frac{1623}{625 \pi^4}-\frac{8}{15}\right) \varrho_5
        +  \frac{21 \left(1000 \pi^4-5769\right) \varrho_7}{6250 \pi^6}+\left(\frac{32}{45}+\frac{21217203}{78125 \pi^8}
        - \frac{30296}{625 \pi^4}\right) \varrho_9$ \\
$(2,5)$ & $-\frac{21 \left(227299527-40383000 \pi^4+500000 \pi^8\right)\varrho_{11}}{781250 \pi^{10}}
        + \left(-\frac{128}{91}+\frac{3938040945531}{19531250 \pi^{12}}-\frac{2800670796}{78125 \pi^8}
        + \frac{285648}{625 \pi^4}\right) \varrho_{13}$\\
        & $+\frac{21 \left(-426619774001421+75842275509000 \pi^4-962461500000 \pi^8
        + 2500000000 \pi^{12}\right) \varrho_{15}}{976562500 \pi^{14}}$ \\
        & $+\left(\frac{512}{153}+\frac{671926478816876283}{1220703125 \pi^{16}}-\frac{955631269448856}{9765625 \pi^{12}}
        + \frac{97089920928}{78125 \pi^8}-\frac{415488}{125 \pi^4}\right) \varrho_{17}$ \\
        & $-\frac{21 \left(122386598885647295199-21757608474072471000 \pi^4+276282575068500000 \pi^8
        - 735547500000000 \pi^{12}+625000000000 \pi^{16}\right) \varrho_{19}}{61035156250 \pi^{18}} + \dots$  \\
\hline
$(4,4)$ & $2\varrho_1+\left(\frac{3}{16 \pi^4}-\frac{8}{15}\right) \varrho_5+\left(\frac{32}{45}+\frac{315}{256 \pi^8}
        - \frac{7}{2 \pi^4}\right) \varrho_9
        + \left(-\frac{128}{91}+\frac{467775}{8192 \pi^{12}}-\frac{10395}{64 \pi^8}+\frac{33}{\pi^4}\right) \varrho_{13}$ \\
       & $+\left(\frac{512}{153}+\frac{638512875}{65536 \pi^{16}}-\frac{14189175}{512 \pi^{12}}+\frac{45045}{8 \pi^8}
       - \frac{240}{\pi^4}\right) \varrho_{17}+ \dots$   \\
	\hline 
\end{tabular}
\caption{First order perturbative corrections (divided by the corresponding energy in the square) for the first twenty  
states of a square drum subject to an arbirtrary (weak) deformation. For the states marked with $\dagger$ the degeneracy 
is lifted by the perturbation.}
\label{table-4}
\end{table}

\end{widetext}

\begin{widetext}
\begin{center}
\begin{table}[htb]

\begin{tabular}{|c||c|c|c||c|c|c|}
	\hline
        & \multicolumn{3}{|c||}{$\lambda=1/100$}  & \multicolumn{3}{|c|}{$\lambda=1/20$}  \\
\hline
$(k,n)$ & $PT_0$ & $PT_2$ &  CCM  & $PT_0$ & $PT_2$ &  CCM   \\
\hline \hline
$(0, 1)$  & 5.78434  &   5.78319  &   5.78319  &   5.81210  &   5.78304  &   5.78325\\
$(1, 1)^\dagger$  & 14.6849  &   14.6805  &   14.6805  &   14.7554  &   14.6447  &   14.646\\
$(1, 1)^\dagger$  & 14.6849  &   14.6834  &   14.6834  &   14.7554  &   14.7185  &   14.7183\\
$(2, 1)$  & 26.3799  &   26.3746  &   26.3746  &   26.5065  &   26.3740  &   26.3734\\
$(2, 1)$  & 26.3799  &   26.3746  &   26.3746  &   26.5065  &   26.3740  &   26.3741\\
$(0, 2)$  & 30.4774  &   30.4713  &   30.4713  &   30.6236  &   30.4705  &   30.4739\\
$(3, 1)$  & 40.7146  &   40.7065  &   40.7065  &   40.9100  &   40.7054  &   40.7056\\
$(3, 1)$  & 40.7146  &   40.7065  &   40.7065  &   40.9100  &   40.7054  &   40.7056\\
$(1, 2)$  & 49.2283  &   49.2135  &   49.2136  &   49.4645  &   49.0936  &   49.0991\\
$(1, 2)$  & 49.2283  &   49.2234  &   49.2234  &   49.4645  &   49.3409  &   49.3415\\
$(4, 1)$  & 57.5945  &   57.5829  &   57.5830  &   57.8709  &   57.5815  &   57.582\\
$(4, 1)$  & 57.5945  &   57.5829  &   57.5830  &   57.8709  &   57.5815  &   57.582\\
$(2, 2)$  & 70.8642  &   70.8500  &   70.8500  &   71.2042  &   70.8482  &   70.8389\\
$(2, 2)$  & 70.8642  &   70.8500  &   70.8500  &   71.2042  &   70.8482  &   70.8501\\
$(0, 3)$  & 74.9020  &   74.8870  &   74.8871  &   75.2614  &   74.8851  &   74.9035\\
$(5, 1)$  & 76.9543  &   76.9389  &   76.9390  &   77.3236  &   76.9370  &   76.9378\\
$(5, 1)$  & 76.9543  &   76.9389  &   76.9390  &   77.3236  &   76.9370  &   76.9378\\
$(3, 2)$  & 95.2966  &   95.2776  &   95.2776  &   95.7540  &   95.2752  &   95.2734\\
$(3, 2)$  & 95.2966  &   95.2776  &   95.2776  &   95.7540  &   95.2752  &   95.2736\\
$(6, 1)$  & 98.7460  &   98.7263  &   98.7263  &   99.2199  &   98.7238  &   98.7251\\
$(6, 1)$  & 98.7460  &   98.7263  &   98.7263  &   99.2199  &   98.7238  &   98.7251\\
$(1, 3)^\dagger$  & 103.520  &   103.489  &   103.489  &   104.017  &   103.237  &   103.253\\
$(1, 3)^\dagger$  & 103.520  &   103.510  &   103.510  &   104.017  &   103.757  &   103.763\\
$(4, 2)$  & 122.452  &   122.428  &   122.428  &   123.040  &   122.425  &   122.422\\
$(4, 2)$  & 122.452  &   122.428  &   122.428  &   123.040  &   122.425  &   122.422\\
$(7, 1)$  & 122.932  &   122.908  &   122.908  &   123.522  &   122.904  &   122.908\\
$(7, 1)$  & 122.932  &   122.908  &   122.908  &   123.522  &   122.904  &   122.908\\
$(2, 3)$  & 135.048  &   135.021  &   135.021  &   135.696  &   135.017  &   134.978\\
$(2, 3)$  & 135.048  &   135.021  &   135.021  &   135.696  &   135.017  &   135.025\\
$(0, 4)$  & 139.068  &   139.040  &   139.041  &   139.735  &   139.037  &   139.098\\
$(8, 1)$  & 149.483  &   149.453  &   149.453  &   150.200  &   149.449  &   149.451\\
$(8, 1)$  & 149.483  &   149.453  &   149.453  &   150.200  &   149.449  &   149.451\\
$(5, 2)$  & 152.272  &   152.241  &   152.241  &   153.002  &   152.237  &   152.238\\
$(5, 2)$  & 152.272  &   152.241  &   152.241  &   153.002  &   152.237  &   152.238\\
$(3, 3)$  & 169.429  &   169.395  &   169.396  &   170.242  &   169.391  &   169.383\\
$(3, 3)$  & 169.429  &   169.395  &   169.396  &   170.242  &   169.391  &   169.384\\
$(1, 4)^\dagger$  & 177.556  &   177.503  &   177.503  &   178.408  &   177.070  &   177.108\\
$(1, 4)^\dagger$  & 177.556  &   177.539  &   177.539  &   178.408  &   177.962  &   177.982\\
$(9, 1)$  & 178.373  &   178.337  &   178.338  &   179.229  &   178.333  &   178.335\\
$(9, 1)$  & 178.373  &   178.337  &   178.338  &   179.229  &   178.333  &   178.335\\
	\hline 
\end{tabular}
\caption{First $40$ eigenvalues of the Robnik billiard with $\lambda = 1/100$ and $\lambda=1/20$. 
The CCM results are obtained with a grid with $N=100$. The symbol $\dagger$ 
is used to highlight the states where first order perturbation theory predicts a lifting  of the degeneracy.}
\label{table-5}
\end{table}
\end{center}
\end{widetext}

\begin{widetext}
\begin{center}
\begin{table}[htb]
\begin{tabular}{|c||c|c|c||c|c|c||c|c|c|}
\hline
        & \multicolumn{3}{|c||}{octagon}  & \multicolumn{3}{|c||}{nonagon} & \multicolumn{3}{|c|}{decagon} \\
\hline 
$(k,n)$ & $PT_0$ & $PT_1$ & CCM  &  $PT_0$ & $PT_1$ & CCM  &   $PT_0$ & $PT_1$ & CCM  \\
\hline \hline
$(0,1)$ & 6.48669  &   6.48505  &   6.48493 & 6.32407  &   6.32314  &   6.32309 & 6.21258   &    6.21202   &    6.21200 \\
$(1,1)$ & 16.468   &   16.4581  &   16.4561 & 16.0551  &   16.0495  &   16.0486 & 15.7721   &    15.7687   &    15.7682 \\
$(1,1)$ & 16.468   &   16.4581  &   16.4561 & 16.0551  &   16.0495  &   16.0486 & 15.7721   &    15.7687   &    15.7682 \\
$(2,1)$ & 29.583   &   29.5530  &   29.5406 & 28.8413  &   28.8241  &   28.8185 & 28.3329   &    28.3224   &    28.3197 \\
$(2,1)$ & 29.583   &   29.5530  &   29.5406 & 28.8413  &   28.8241  &   28.8186 & 28.3329   &    28.3224   &    28.3197 \\
$(0,2)$ & 34.178   &   34.1407  &   34.1245 & 33.3211  &   33.2994  &   33.2920 & 32.7337   &    32.7204   &    32.7166 \\
$(3,1)$ & 45.6583  &   45.5912  &   45.5298 & 44.5136  &   44.4747  &   44.4501 & 43.7289   &    43.7050   &    43.6936 \\
$(3,1)$ & 45.6583  &   45.5912  &   45.5298 & 44.5136  &   44.4747  &   44.4501 & 43.7289   &    43.7050   &    43.6937 \\
$(1,2)$ & 55.2057  &   55.1197  &   55.0498 & 53.8217  &   53.7708  &   53.7391 & 52.8729   &    52.8412   &    52.8254 \\
$(1,2)$ & 55.2057  &   55.1197  &   55.0498 & 53.8217  &   53.7708  &   53.7391 & 52.8729   &    52.8412   &    52.8255 \\
$(4,1)$ & 64.5877  &   62.6348  &   62.5959$^\dagger$ & 62.9685  &   62.8946  &   62.7662 & 61.8584   &    61.8128   &    61.7682 \\
$(4,1)$ & 64.5877  &   66.2878  &   66.2775$^\dagger$ & 62.9685  &   62.8946  &   62.7662 & 61.8584   &    61.8128   &    61.7682 \\
$(2,2)$ & 79.4687  &   79.3097  &   79.0194 & 77.4763  &   77.3810  &   77.2694 & 76.1106   &    76.0505   &    75.9987 \\
$(2,2)$ & 79.4687  &   79.3097  &   79.0199 & 77.4763  &   77.3810  &   77.2696 & 76.1106   &    76.0505   &    75.9988 \\
$(0,3)$ & 83.9968  &   83.8282  &   83.5906 & 81.8909  &   81.7892  &   81.6829 & 80.4473   &    80.3828   &    80.3302 \\
$(5,1)$ & 86.2983  &   86.0854  &   85.8297 & 84.1347  &   84.0094  &   84.0635 & 82.6516   &    81.0526   &    81.0237$^\dagger$  \\
$(5,1)$ & 86.2983  &   86.0854  &   85.8297 & 84.1347  &   84.0094  &   84.0635 & 82.6516   &    84.0950   &    84.0826$^\dagger$  \\
$(3,2)$ & 106.868  &   106.609  &   106.713 & 104.189  &   104.032  &   102.757 & 102.352   &    102.252   &    102.055 \\
$(3,2)$ & 106.868  &   106.609  &   106.713 & 104.189  &   104.032  &   102.757 & 102.352   &    102.252   &    102.055 \\
$(6,1)$ & 110.736  &   110.405  &   110.452 & 107.960  &   107.763  &   108.953 & 106.057   &    105.934   &    105.808 \\
$(6,1)$ & 110.736  &   110.405  &   110.452 & 107.960  &   107.763  &   108.953 & 106.057   &    105.934   &    105.808 \\
$(1,3)$ & 116.090  &   115.812  &   114.883 & 113.179  &   113.009  &   112.684 & 111.184   &    111.075   &    110.926 \\
$(1,3)$ & 116.090  &   115.812  &   114.883 & 113.179  &   113.009  &   112.684 & 111.184   &    111.075   &    110.927 \\
$(4,2)$ & 137.321  &   133.003  &   132.733$^\dagger$ & 133.878  &   133.641  &   133.313 & 131.518   &    131.366   &    131.401 \\
$(4,2)$ & 137.321  &   137.373  &   137.954$^\dagger$ & 133.878  &   133.641  &   133.313 & 131.518   &    131.366   &    131.401 \\
$(7,1)$ & 137.859  &   137.373  &   137.954 & 134.403  &   134.113  &   133.485 & 132.033   &    131.851   &    131.925 \\
$(7,1)$ & 137.859  &   140.865  &   140.802 & 134.403  &   134.113  &   133.485 & 132.033   &    131.851   &    131.926 \\
$(2,3)$ & 151.446  &   151.031  &   150.459 & 147.649  &   147.391  &   147.814 & 145.046   &    144.879   &    144.205 \\
$(2,3)$ & 151.446  &   151.031  &   150.461 & 147.649  &   147.391  &   147.815 & 145.046   &    144.879   &    144.205 \\
$(0,4)$ & 155.954  &   155.530  &   151.944 & 152.044  &   151.780  &   150.877 & 149.364   &    149.192   &    148.819 \\
$(8,1)$ & 167.633  &   165.334  &   168.290$^\dagger$ & 163.431  &   163.022  &   162.822  & 160.550   &    160.292   &    160.168 \\
$(8,1)$ & 167.633  &   168.569  &   168.776$^\dagger$ & 163.431  &   163.022  &   162.822 & 160.550   &    160.292   &    160.817 \\
$(5,2)$ & 170.761  &   170.216  &   168.776 & 166.480  &   166.143  &   166.165 & 163.545   &    160.293   &    160.817$^\dagger$  \\
$(5,2)$ & 170.761  &   170.216  &   169.209 & 166.480  &   166.143  &   166.165 & 163.545   &    166.360   &    166.328$^\dagger$  \\
$(3,3)$ & 190.002  &   189.422  &   189.910 & 185.238  &   184.875  &   183.416 & 181.973   &    181.735   &    181.436 \\
$(3,3)$ & 190.002  &   189.422  &   189.910 & 185.238  &   184.875  &   183.416 & 181.973   &    181.735   &    181.437 \\
$(1,4)$ & 199.116  &   198.519  &   192.516 & 194.124  &   193.008  &   193.117 & 190.702   &    190.454   &    186.961 \\
$(1,4)$ & 199.116  &   198.519  &   192.516 & 194.124  &   193.748  &   193.117 & 190.702   &    190.454   &    186.961 \\
$(9,1)$ & 200.032  &   199.109  &   204.554 & 195.017  &   193.748  &   193.659$^\dagger$ & 191.579   &    191.226   &    194.384 \\
$(9,1)$ & 200.032  &   199.109  &   204.554 & 195.017  &   195.913  &   195.691$^\dagger$ & 191.579   &    191.226   &    194.384 \\
	\hline 
\end{tabular}
\caption{First $40$ eigenvalues of the octagon, nonagon and decagon billiards. The symbol $\dagger$ 
is used to highlight the states where first order perturbation theory predicts a lifting  of the degeneracy.
The third column for each polygon displays the numerical result obtained using the CCM with a grid corresponding
to $N=100$.}
\label{table-6}
\end{table}
\end{center}
\end{widetext}

\end{document}